\documentclass[journal]{IEEEtran}
\usepackage{amssymb}
\usepackage{algorithmic}
\usepackage{algorithm}
\usepackage{amsfonts}
\usepackage{amsmath}
\usepackage{amssymb}
\usepackage{amsthm}
\usepackage{array}
\usepackage{bbding}
\usepackage{bigints}
\usepackage{booktabs}
\usepackage{cite}
\usepackage{cleveref}
\usepackage{color}
\usepackage{diagbox}
\usepackage{epsfig,latexsym}
\usepackage{epstopdf}
\usepackage{graphicx}
\usepackage{fancyhdr}
\usepackage{float}
\usepackage{flushend}
\usepackage{indentfirst}
\usepackage{lastpage}
\usepackage{makecell}
\usepackage{mathtools}
\usepackage{multirow}
\usepackage{pifont}
\usepackage{psfrag}
\usepackage{setspace}
\usepackage{stfloats}
\usepackage{subfloat}
\usepackage{tikz}
\usepackage{times}
\usepackage{subfig}

\theoremstyle{remark}

\allowdisplaybreaks[4]

\begin{document}
\title{Energy-Efficient Reconfigurable Holographic Surfaces Operating in the Presence of Realistic Hardware Impairments}

\author{Qingchao Li, \textit{Graduate Student Member, IEEE}, Mohammed El-Hajjar, \textit{Senior Member, IEEE},\\ Yanshi Sun, \textit{Member, IEEE}, Ibrahim Hemadeh, \textit{Senior Member, IEEE},\\ Arman Shojaeifard, \textit{Senior Member, IEEE}, and Lajos Hanzo, \textit{Life Fellow, IEEE}

\thanks{Lajos Hanzo would like to acknowledge the financial support of the Engineering and Physical Sciences Research Council projects EP/W016605/1, EP/X01228X/1, EP/Y026721/1 and EP/W032635/1 as well as of the European Research Council's Advanced Fellow Grant QuantCom (Grant No. 789028). \textit{(Corresponding author: Lajos Hanzo.)}

Qingchao Li, Mohammed El-Hajjar and Lajos Hanzo are with the School of Electronics and Computer Science, University of Southampton, Southampton SO17 1BJ, U.K. (e-mail: qingchao.li@soton.ac.uk; meh@ecs.soton.ac.uk; lh@ecs.soton.ac.uk).

Yanshi Sun is with the School of Computer Science and Information Engineering, Hefei University of Technology, Hefei, 230009, China. (email:sys@hfut.edu.cn).

Ibrahim Hemadeh and Arman Shojaeifard are with InterDigital, London EC2A 3QR, U.K. (e-mail: ibrahim.hemadeh@interdigital.com; arman.shojaeifard@interdigital.com).}}

\maketitle

\begin{abstract}
Reconfigurable holographic surfaces (RHSs) constitute a promising technique of supporting energy-efficient communications. In this paper, we formulate the energy efficiency maximization problem of the switch-controlled RHS-aided beamforming architecture by alternately optimizing the holographic beamformer at the RHS, the digital beamformer, the total transmit power and the power sharing ratio of each user. Specifically, to deal with this challenging non-convex optimization problem, we decouple it into three sub-problems. Firstly, the coefficients of RHS elements responsible for the holographic beamformer are optimized to maximize the sum of the eigen-channel gains of all users by our proposed low-complexity eigen-decomposition (ED) method. Then, the digital beamformer is designed by the singular value decomposition (SVD) method to support multi-user information transfer. Finally, the total transmit power and the power sharing ratio are alternately optimized, while considering the effect of transceiver hardware impairments (HWI). We theoretically derive the spectral efficiency and energy efficiency performance upper bound for the RHS-based beamforming architectures in the presence of HWIs. Our simulation results show that the switch-controlled RHS-aided beamforming architecture achieves higher energy efficiency than the conventional fully digital beamformer and the hybrid beamformer based on phase shift arrays (PSA). Moreover, considering the effect of HWI in the beamforming design can bring about further energy efficiency enhancements.
\end{abstract}
\begin{IEEEkeywords}
Reconfigurable holographic surfaces (RHS), hybrid beamforming, energy efficiency, eigen-decomposition (ED), hardware impairment (HWI).
\end{IEEEkeywords}

\section{Introduction}
\IEEEPARstart{D}{riven} by the ever-increasing demand for data sharing across mobile networks, sophisticated technologies, such as millimeter wave (mmWave) communications~\cite{zeng2016millimeter}, \cite{tang2022path}, massive multiple-input-multiple-output (MIMO) systems~\cite{ma2018sparse}, \cite{he2022beamspace}, \cite{du2022tensor}, \cite{li2023uav}, and ultra-dense networks (UDN)~\cite{pan2018robust}, \cite{pan2018joint} have been rolled out across the globe. Hence, at the time of writing research is also well under way for the exploration of next-generation communications~\cite{huang2022general}.

In this context, holographic MIMO systems are expected to evolve toward an intelligent and software reconfigurable paradigm enabling ubiquitous communications between humans and mobile devices in support of improved spectral efficiency~\cite{huang2020holographic}.

From the perspective of holographic MIMO channel models, Pizzo \textit{et al.}~\cite{pizzo2020spatially} modelled the far-field small-scale fading of holographic MIMO systems as a zero-mean, spatially-stationary, and correlated Gaussian scalar random field, where a Fourier plane-wave based spectral representation is employed to describe the three-dimensional small-scale fading. Furthermore, the near-field small-scale fading of holographic MIMO systems is derived in~\cite{pizzo2022fourier}, where a novel Fourier plane-wave stochastic channel model is conceived, which is mathematically tractable and consistent with the physics of wave propagation. In~\cite{demir2022channel}, Demir \textit{et al.} proposed a spatial correlation model to characterize the correlated Rayleigh fading channel, considering both non-isotropic scattering and directive antennas. In~\cite{zeng2022reconfigurable}, Zeng \textit{et al.} considered a single-user holographic MIMO system model, where a reconfigurable refractive surface is deployed at the BS empowered by a single feed, and its downlink data rate is derived.

As a further advance, a reconfigurable holographic surface (RHS)-aided beamforming architecture is proposed in~\cite{deng2021reconfigurable_review}, \cite{deng2023reconfigurable}. Specifically, the RHS is composed of feeds and metamaterial based radiation elements, in which the feeds connect the radio frequency (RF) chains and the RHS circuit. These schemes transform the input signals into electromagnetic waves, and the coefficients of the radiation elements are controlled electrically to achieve holographic beamforming. In~\cite{deng2021reconfigurable_tvt}, \cite{deng2022hdma}, \cite{deng2022reconfigurable_wc}, Deng \textit{et al.} proposed an RHS-based hybrid beamforming scheme, where the digital beamformer and the holographic beamformer are realized at the BS and the RHS, respectively. Specifically, the digital beamformer relies on the state-of-the-art zero-forcing (ZF) precoding method, while the holographic beamforming is performed by configuring the amplitude-controlled RHS radiation elements. To maximize the achievable sum-rate, the digital beamformer and the holographic beamformer are alternately optimized. The simulation results showed that the proposed RHS-based hybrid beamformer achieves better sum-rate than the state-of-the-art massive MIMO hybrid beamformer based on phase shift arrays. In~\cite{hu2022holographic}, Hu \textit{et al.} proposed a holographic beamforming scheme based on amplitude-controlled RHS elements having limited resolution, and it showed that the holographic beamforming scheme associated with 2-bit quantized amplitude resolution approached the sum-rate associated with continuous amplitude values. In~\cite{deng2022holographic}, an amplitude-controlled RHS beamformer was harnessed for ultra-dense low-Earth-orbit (LEO) satellite communications to compensate for the severe path-loss of satellite communications.

\textit{The above treatises on RHS-aided systems have the following limitations.} Firstly, the objective function in these works mainly focused on maximizing the data rate, whilst ignoring the optimization of the energy efficiency. However, according to the report from the Radiocommunications Sector of the International Telecommunication Union (ITU-R) M.2516-0, improving the energy efficiency is one of the key drivers for International Mobile Telecommunications (IMT) systems designed for sustainable development. Power-efficient solutions are needed both for the backhaul and for local access to make use of small-scale renewable energy sources. Secondly, perfect transceiver hardware is assumed, i.e. the practical signal impairments of the transceivers are ignored. However, the hardware impairments (HWI) may impose a performance floor at high transmit powers, both in terms of the achievable rate, outage probability and bit error ratio (BER)~\cite{yang2021performance}, \cite{peng2021ris}, \cite{vu2022performance}, \cite{li2023performance}, \cite{wang2023ris}, \cite{li2023achievable}, \cite{li2024achievable}. Hence the attainable performance improvement of wireless communication systems is limited by the HWI, which cannot be compensated upon implying increasing the transmit power. To deal with the above issues, our contributions in this paper can be summarized as follows:

\begin{itemize}
  \item We formulate an energy efficiency maximization problem for the switch-controlled RHS-enabled wireless communication systems operating in the face of realistic transceiver HWIs. More explicitly, we optimize the holographic beamformer, the digital beamformer, the total transmit power, and the power allocation ratio of each user. We then solve the problem by  decomposing it into three sub-problems. Firstly, in the holographic beamformer, the optimal ON-OFF state of each RHS radiation element is found based on our proposed low-complexity eigen-decomposition (ED) method to maximize the sum of the eigen-channel gains of all users. Then, the digital beamformer is designed based on the singular value decomposition (SVD) method to support multi-user information transfer. Finally, the optimal transmit power and the power sharing ratio are alternately optimized. Specifically, when the total transmit power is given, the proportion of power allocated for each user is optimized via our proposed power sharing algorithm, while considering the effect of HWIs. By contrast, when the power share of each user is determined, the total transmit power is optimized based on the gradient descent method.
  \item To investigate the effect of HWI, we theoretically derive the spectral efficiency and energy efficiency upper bound for the RHS-based beamforming architectures in the presence of HWIs.
  \item The simulation results show that the switch-controlled RHS-aided beamforming architecture achieves higher energy efficiency than the conventional digital beamformer and the hybrid beamformer based on a phase shift array. Furthermore, we show that considering the effect of HWIs in the beamforming design yields further energy efficiency enhancements.
\end{itemize}

The rest of this paper is organized as follows. In Section~\ref{System_Model} we present the system model, while our reconfigurable holographic surface-based hybrid beamformer design is described in Section~\ref{System_Model}. Our simulation results are presented in Section~\ref{Numerical_and_Simulation_Results}, while we conclude in Section~\ref{Conclusion}.

\textit{Notations:} Vectors and matrices are denoted by boldface lower and upper case letters, respectively; $(\cdot)^{\text{T}}$, $(\cdot)^{\dag}$ and $(\cdot)^{\text{H}}$ represent the operation of transpose, conjugate and Hermitian transpose, respectively; $|a|$ and $\angle a$ represent the amplitude and angle of the complex scalar $a$, respectively; $\mathbf{1}_{n}$ represent the $n\times1$ one vector; $\mathbb{R}^{m\times n}$ and $\mathbb{C}^{m\times n}$ denotes the space of $m\times n$ real-valued and complex-valued matrices, respectively; $a_n$ represents the $n$th element in vector $\mathbf{a}$, and $[\mathbf{A}]_{m,n}$ represents the $(m,n)$th element in matrix $\mathbf{A}$; $\mathbf{Diag}\{a_1,a_2,\cdots,a_N\}$ denotes a diagonal matrix with the diagonal elements being the elements of $a_1,a_2,\cdots,a_N$ in order; $\mathrm{rank}(\mathbf{A})$ denotes the rank of matrix $\mathbf{A}$; $\mathcal{CN}(\boldsymbol{\mu},\mathbf{\Sigma})$ is a circularly symmetric complex Gaussian random vector with the mean $\boldsymbol{\mu}$ and the covariance matrix $\mathbf{\Sigma}$; $\mathbf{A}\succeq\mathbf{0}$ means $\mathbf{A}$ is a semi-definite matrix; $\mathbf{0}_n$ represents the $n\times1$ zeros vector and $\mathbf{I}_n$ represents the $n\times n$ identical matrix; $\odot$ denotes Hadamard product operation, and $\mathcal{O}$ represents the time complexity.

\section{System Model}\label{System_Model}
In this section, we briefly highlight the energy-efficient switch-controlled RHS-aided beamforming architecture of~\cite{deng2021reconfigurable_review}, \cite{deng2023reconfigurable}, \cite{deng2021reconfigurable_tvt}, \cite{deng2022hdma}, \cite{deng2022reconfigurable_wc}, \cite{hu2022holographic}, \cite{deng2022holographic}, followed by the channel model.

\begin{figure*}[!t]
   \centering
    \includegraphics[width=7in]{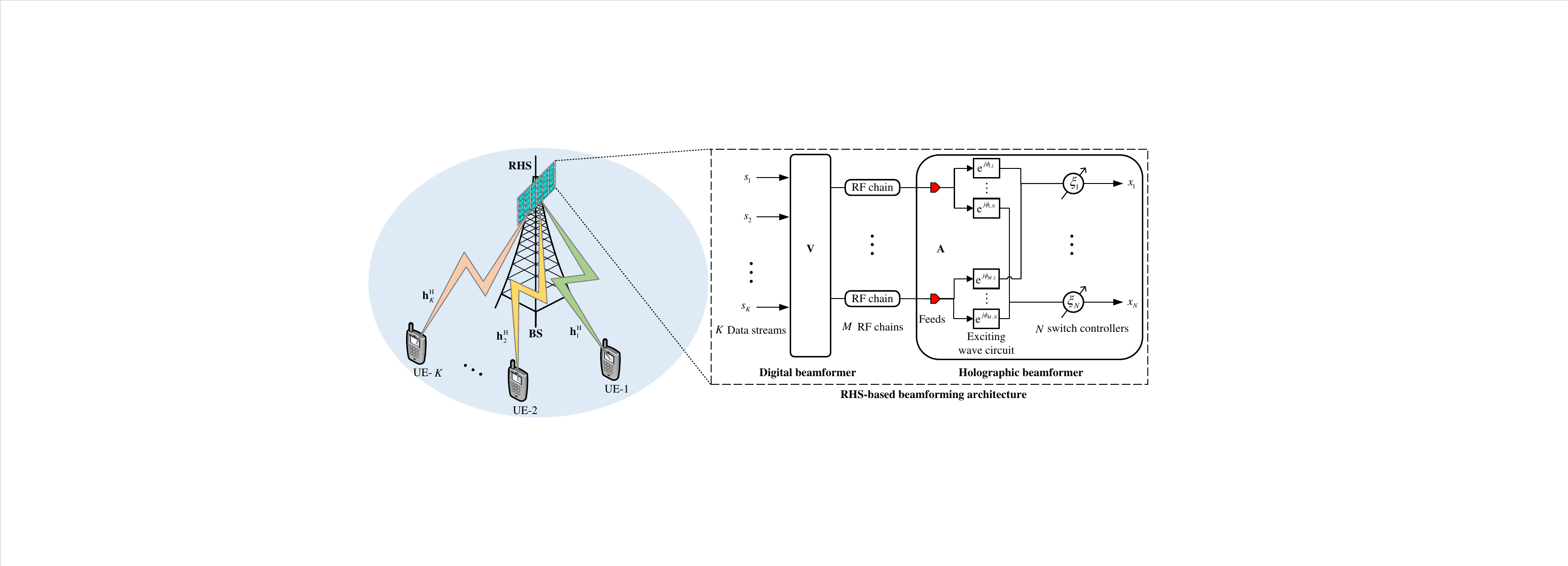}
    \caption{System model of the RHS-based hybrid beamforming architecture.}\label{Fig_RHS_architecture}
\end{figure*}

\subsection{Reconfigurable Holographic Surface aided Beamforming Architecture}\label{RHS_Architecture}
The physical principle of RHS is different from some other low-power aperture antenna designs, such as the dynamic metasurface antennas (DMA)~\cite{you2022energy}, \cite{jiang2022hybrid}, \cite{shlezinger2021dynamic}, \cite{wang2020dynamic}. Specifically, the DMA consists of multiple microstrips and each microstrip is composed of numerous metamaterial elements. In the DMA, the information is beamformed by linearly combining the radiation observation from all metamaterial elements in each microstrip. By contrast, the RHS is a special leaky-wave antenna and the information is beamformed based on an exciting wave circuit and a radiation element configuration controller.

Referring to~\cite{deng2021reconfigurable_review}, \cite{deng2023reconfigurable}, \cite{deng2021reconfigurable_tvt}, \cite{deng2022hdma}, \cite{deng2022reconfigurable_wc}, \cite{hu2022holographic}, \cite{deng2022holographic}, the architecture of the RHS-based hybrid beamformer is shown in Fig.~\ref{Fig_RHS_architecture}, including both the digital and the holographic beamformer.

\subsubsection{Digital beamformer architecture}
Firstly, the data stream vector $\mathbf{s}=[s_1,s_2,\cdots,s_K]^\mathrm{T}\in\mathbb{C}^{K\times1}$ is processed by a power allocation matrix $\mathbf{P}\in\mathbb{C}^{K\times K}$ and a digital beamformer $\mathbf{V}=[\mathbf{v}_1,\mathbf{v}_2,\cdots,\mathbf{v}_K]\in\mathbb{C}^{M\times K}$, where $K$ is the number of data streams. Then, the resultant signal is up-converted to the carrier frequency by $M$ RF chains, where we have $\mathbb{E}[|s_k|^2]=1$, $\mathbf{P}=\mathbf{Diag}\{p_1,p_2,\cdots,p_K\}$ satisfying $\sum_{k=1}^{K}p_k=1$, and $\|\mathbf{v}_k\|^2=1$.

\subsubsection{Holographic beamformer architecture}
In the RHS-based holographic beamformer, each RF chain is connected to a feed of the RHS, and the feeds convert the carrier frequency current into an electromagnetic wave, which propagates through the waveguide of the reconfigurable holographic surface and radiates the energy into the free space from the metamaterial radiation elements. Specifically, at the $n$th RHS element, the desired wave propagates in the target direction characterized by $\Psi_\mathrm{obj}(\mathbf{r}_n)=\mathrm{e}^{-j\mathbf{k}_n^{\mathrm{T}}\cdot\mathbf{r}_n}$, while the reference wave generated by the $m$th feed is $\Psi_\mathrm{ref}(\mathbf{r}_n^m)=\mathrm{e}^{-j\mathbf{k}_0^{\mathrm{T}}\cdot\mathbf{r}_n^m}$, where $\mathbf{k}_n$ is the desired directional propagation vector emerging from the $n$th RHS element, $\mathbf{k}_0$ is the propagation vector of the reference wave, $\mathbf{r}_n$ is the position vector of the $n$th RHS element, and $\mathbf{r}_n^m$ is the position vector representing the link spanning from the $m$th feed to the $n$th RHS element. As shown in Fig.~\ref{Fig_RHS_architecture}, the holographic beamformer matrix, denoted as $\mathbf{A}$, is composed of the fixed exciting wave circuit matrix $\mathbf{\Phi}\in\mathbb{C}^{N\times M}$ and the diagonal reconfigurable controller matrix $\mathbf{F}\in\mathbb{C}^{N\times N}$, which can be represented as~\cite{deng2021reconfigurable_review}, \cite{deng2023reconfigurable}, \cite{deng2021reconfigurable_tvt}, \cite{deng2022hdma}, \cite{deng2022reconfigurable_wc}, \cite{hu2022holographic}, \cite{deng2022holographic}:
\begin{align}
    \mathbf{A}=\mathbf{F}\mathbf{\Phi}.
\end{align}

In the exciting wave circuit, $[\mathbf{\Phi}]_{n,m}=\frac{1}{\sqrt{N}}\mathrm{e}^{j\phi_{n,m}}$ represents the exciting wave emanating from the $m$th feed to the $n$th RHS element. The holographic pattern in~\cite{deng2021reconfigurable_review}, \cite{deng2023reconfigurable}, \cite{deng2021reconfigurable_tvt}, \cite{deng2022hdma}, \cite{deng2022reconfigurable_wc}, \cite{hu2022holographic}, \cite{deng2022holographic} is excited by the reference wave, i.e. $[\mathbf{\Phi}]_{n,m}=\frac{1}{\sqrt{N}}\mathrm{e}^{-j\mathbf{k}_0^{\mathrm{T}}
\cdot\mathbf{r}_n^m}$. To avoid the impact of the feed and RHS element location on the beamforming, we design the exciting wave circuit based on the discrete Fourier transform (DFT) matrix as follows. In the exciting wave circuit between the $m$th feed and the $n$th RHS element, we apply a calibration phase shifter of $\Delta\phi_{n,m}=\frac{(n-1)(m-1)}{N}+\mathbf{k}_0^{\mathrm{T}}\cdot\mathbf{r}_n^m$ to the reference wave, and then the generated exciting wave is given by
\begin{align}
    [\mathbf{\Phi}]_{n,m}
    =\frac{1}{\sqrt{N}}\mathrm{e}^{j\left(-\mathbf{k}_0^{\mathrm{T}}
    \cdot\mathbf{r}_n^m+\Delta\phi_{n,m}\right)}
    =\frac{1}{\sqrt{N}}\mathrm{e}^{j\frac{(n-1)(m-1)}{N}}.
\end{align}

Additionally, we denote the diagonal reconfigurable controller matrix as $\mathbf{F}=\mathbf{Diag}\{\xi_1,\xi_2,\cdots,\xi_N\}$, where $\xi_n$ is the holographic pattern of the $n$th RHS element. We refer to~\cite{deng2021reconfigurable_review}, \cite{deng2023reconfigurable}, \cite{deng2021reconfigurable_tvt}, \cite{deng2022hdma}, \cite{deng2022reconfigurable_wc}, \cite{hu2022holographic}, \cite{deng2022holographic} to configure the coefficients of each RHS element, where the RHS utilizes an amplitude-controlled method appropriately adjusting the radiation amplitude of the reference wave at each radiation element instead of conventional phase shifters. Specifically, the real part of the holographic pattern, i.e. $\mathrm{Re}[\Psi_\mathrm{intf}(\mathbf{r}_n^m)]$, is the cosine value of the phase difference between the desired object wave $\Psi_\mathrm{obj}(\mathbf{r}_n)$ and the reference wave $\Psi_\mathrm{ref}(\mathbf{r}_n^m)$. Thus, $\mathrm{Re}[\Psi_\mathrm{intf}(\mathbf{r}_n^m)]$ represents the amplitude of each radiation element. To avoid a negative value, $\mathrm{Re}[\Psi_\mathrm{intf}(\mathbf{r}_n^m)]$ is normalized to $[0,1]$, and the coefficient of the $n$th RHS element is given by
\begin{align}
    \xi_n=\frac{\mathrm{Re}\left[\Psi_\mathrm{intf}(\mathbf{r}_n^m)\right]+1}{2},
\end{align}
where we have $\xi_n\in\mathbb{R}$ satisfying $0\leq\xi_n\leq1$. To reduce the power consumption, a switch-controlled method can be employed, which may be realized by PIN diodes, since PIN diodes have two states, i.e. ON and OFF~\cite{deng2022reconfigurable_wc}. Therefore, the diagonal reconfigurable controller matrix is designed as
\begin{align}
    \mathbf{F}=\mathbf{Diag}\left\{\xi_1,\xi_2,\cdots,\xi_N\right\},\quad \xi_n\in\{0,1\}.
\end{align}

\subsection{Channel Model}
As shown in Fig.~\ref{Fig_RHS_architecture}, a base station (BS) relying on RHS-based hybrid beamforming supports $K$ single-antenna user equipment (UE). We denote the large-scale fading and the link spanning from the metamaterial radiation elements to UE-$k$ as $\upsilon_k$ and $\mathbf{h}_k^{\mathrm{H}}\in\mathbb{C}^{1\times N}$, respectively. We adopt the mmWave channel model to characterize the propagation environment between the BS and each UE. Specifically, the mmWave downlink channel of UE-$k$ is assumed to be the sum of all propagation
paths that are scattered in $L_{C}$ clusters and each cluster contributes $L_{P}$ paths, which can be expressed as
\begin{align}
    \mathbf{h}_k^\mathrm{H}=\sqrt{\frac{1}{L_CL_P}}\sum_{c=1}^{L_C}\sum_{p=1}^{L_P}\alpha_{c,p}^k
    \mathbf{a}^\mathrm{H}(\psi_{c,p}^k,\varphi_{c,p}^k),
\end{align}
where $\alpha_{c,p}^k$ is the complex gain of the $p$th path in the $c$th cluster following $\alpha_{c,p}^k\sim\mathcal{CN}(0,1)$, and $\mathbf{a}^\mathrm{H}(\psi_{c,p}^k,\varphi_{c,p}^k)$ is given by
\begin{align}
    \notag&\mathbf{a}^\mathrm{H}(\psi_{c,p}^k,\varphi_{c,p}^k)\\
    \notag=&\left[1,\cdots,\mathrm{e}^{-j\frac{2\pi}{\lambda}\left(\delta_xn_x\sin\psi_{c,p}^k
    \cos\varphi_{c,p}^k+\delta_yn_y\sin\psi_{c,p}^k\sin\varphi_{c,p}^k\right)},\cdots,\right.\\
    &\left.\mathrm{e}^{-j\frac{2\pi}{\lambda}\left(\delta_x\left(N_x-1\right)
    \sin\psi_{c,p}^k\cos\varphi_{c,p}^k
    +\delta_y\left(N_y-1\right)\sin\psi_{c,p}^k\sin\varphi_{c,p}^k\right)}\right],
\end{align}
where $n_x=0,1,\cdots,N_x-1$, $n_y=0,1,\cdots,N_y-1$, $\lambda$ is the wave length of the carrier, $\delta_x$ and $\delta_y$ are the distance of the adjacent RHS radiation elements in the horizontal direction and vertical direction, respectively. Furthermore, $\psi_{c,p}^k$ and $\varphi_{c,p}^k$ are the elevation and azimuth angles of departure (AoD) from the RHS to UE-$k$ in the $p$th path of the $c$th cluster, respectively. Within the $c$th cluster, the random variables $\psi_{c,p}^k$ and $\varphi_{c,p}^k$ have the uniformly distributed mean values of $\mu_{\psi_{c}^k}$ and $\mu_{\varphi_{c}^k}$, respectively, and have the angular spreads, i.e. standard deviations, of $\sigma_{\psi_{c}^k}$ and $\sigma_{\varphi_{c}^k}$, respectively. Furthermore, we define $\mathbf{\Upsilon}=\mathbf{Diag}\{\upsilon_1,\upsilon_2,\cdots,\upsilon_K\}$ and $\mathbf{H}=[\mathbf{h}_1,\mathbf{h}_2,\cdots,\mathbf{h}_K]^\mathrm{H}$. At the receiver, we denote the receiver combining vector at UE-$k$ by $\mathbf{u}_k^\mathrm{H}$ satisfying $\|\mathbf{u}_k\|^2=1$, and define $\mathbf{U}=[\mathbf{u}_1,\mathbf{u}_2,\cdots,\mathbf{u}_k]^\mathrm{H}\in\mathbb{C}^{K\times K}$. Although in multi-user systems typically each UE recovers its information independently, here we assume that the UEs can jointly recover their information to unveil the performance limits of the RHS-based multi-user systems. The joint multi-user detection scheme is also widely employed for the successive interference cancellation (SIC) operation of non-orthogonal multiple access (NOMA) systems~\cite{wang2021minimum}, \cite{de2020user}, \cite{dai2018survey}, where the information recovery of each user may rely on the detection of other users, albeit it increases the channel estimation feedback overheads. Hence, the receiver combining vector $\mathbf{u}_k^\mathrm{H}$ at UE-$k$ is designed based on the signals received from all UEs.

We denote the signals propagating from the metamaterial radiation elements by $\mathbf{x}=[x_1,x_2,\cdots,x_N]^\mathrm{T}\in\mathbb{C}^{N\times1}$, which can be represented as
\begin{align}
    \notag\mathbf{x}=&\sqrt{\rho\varepsilon_\mathrm{t}}\sum_{k=1}^{K}\sqrt{p_k}
    \mathbf{F}\mathbf{\Phi}\mathbf{v}_ks_k\\
    &+\sqrt{\rho\left(1-\varepsilon_\mathrm{t}\right)}
    \sum_{k=1}^{K}\sqrt{p_k}\mathbf{F}\mathbf{\Phi}\mathbf{v}_k\odot\boldsymbol{\eta}_k,
\end{align}
where $\rho$ denotes the total transmit power at the BS. Furthermore, $\boldsymbol{\eta}_k\sim\mathcal{CN}(\mathbf{0}_M,\mathbf{I}_M)$ is the distortion of the information symbol $s_k$ due to hardware impairments at the RF chains, resulting from power amplifier non-linearities, amplitude/phase imbalance in the In-phase/Quadrature mixers, phase noise in the local oscillator, sampling jitter and finite-resolution quantization in the analog-to-digital converters~\cite{bjornson2017massive}, at the BS. Finally, $\varepsilon_{\mathrm{t}}$ is the hardware quality factor of the RF-chains at the BS satisfying $0\leq\varepsilon_{\mathrm{t}}\leq1$. Explicitly, a hardware quality factor of 1 indicates that the hardware is ideal, while 0 means that the hardware is completely inadequate. Therefore, the signal received by UE-$k$ is given in (\ref{Channel_Model_1}),
\begin{figure*}[!t]
\begin{align}\label{Channel_Model_1}
    \notag y_k=&\underbrace{\sqrt{\rho\varepsilon_\mathrm{r}\varepsilon_\mathrm{t}p_k\upsilon_k}
    \mathbf{u}_k^{\mathrm{H}}\mathbf{H}\mathbf{F}\mathbf{\Phi}\mathbf{v}_ks_k}
    _{\text{Desired signal of the UE-$k$ information}}
    +\underbrace{\sqrt{\rho\varepsilon_\mathrm{r}\left(1-\varepsilon_\mathrm{t}\right)p_k\upsilon_k}
    \mathbf{u}_k^{\mathrm{H}}\mathbf{H}\mathbf{F}\mathbf{\Phi}
    \left(\mathbf{v}_k\odot\boldsymbol{\eta}_k\right)}
    _{\text{BS HWI distortion on the UE-$k$ information}}
    +\underbrace{\sqrt{\rho\left(1-\varepsilon_\mathrm{r}\right)p_k\upsilon_k}
    \mathbf{u}_k^{\mathrm{H}}\mathbf{H}\mathbf{F}\mathbf{\Phi}\mathbf{v}_k\varsigma_k}
    _{\text{UE-$k$ HWI distortion on the UE-$k$ information}}+\\
    \notag&\underbrace{\sum_{k'\neq k}
    \left(\sqrt{\rho\varepsilon_\mathrm{r}\varepsilon_\mathrm{t}p_{k'}\upsilon_k}
    \mathbf{u}_k^{\mathrm{H}}\mathbf{H}\mathbf{F}\mathbf{\Phi}\mathbf{v}_{k'}s_{k'}
    +\sqrt{\rho\varepsilon_\mathrm{r}\left(1-\varepsilon_\mathrm{t}\right)p_{k'}\upsilon_k}
    \mathbf{u}_k^{\mathrm{H}}\mathbf{H}\mathbf{F}\mathbf{\Phi}
    \left(\mathbf{v}_{k'}\odot\boldsymbol{\eta}_{k'}\right)
    +\sqrt{\rho\left(1-\varepsilon_\mathrm{r}\right)p_{k'}\upsilon_k}
    \mathbf{u}_k^{\mathrm{H}}\mathbf{H}\mathbf{F}\mathbf{\Phi}\mathbf{v}_{k'}\varsigma_{k'}\right)}
    _{\text{Inter-user interference}}\\
    &+\underbrace{w_k}_{\text{Additive noise}},
\end{align}
\hrulefill
\end{figure*}
where $w_k\sim\mathcal{CN}(0,\sigma_w^2)$ represents the additive noise at UE-$k$. Furthermore, $\boldsymbol{\varsigma}=[\varsigma_1,\varsigma_2,\cdots,\varsigma_K]^{\mathrm{T}}$, with $\varsigma_k\sim\mathcal{CN}(0,1)$ representing the distortion of the information symbol $s_k$ due to the hardware impairments at UE-$k$, and the hardware quality factor of the UEs satisfying $0\leq\varepsilon_\mathrm{r}\leq1$.

\section{Reconfigurable Holographic Surface Based Hybrid Beamformer Design}\label{Hybrid_Beamforming_Design}
In this section, we formulate the energy efficiency maximization problem of the switch-controlled RHS beamformer. Then, we decouple it into three sub-problems, including the holographic beamforming, digital beamforming, and power sharing ratio and total transmit power optimization.

According to (\ref{Channel_Model_1}), the received signal-to-interference-plus-noise-ratio (SINR) of the information $s_k$ at UE-$k$ is given by
\begin{align}\label{Hybrid_Beamforming_Design_1}
    \gamma_k=\frac{\rho\varepsilon_\mathrm{r}\varepsilon_\mathrm{t}p_k
    \mathcal{G}_{k,k}}{\rho\left(1-\varepsilon_\mathrm{r}\varepsilon_\mathrm{t}\right)p_k
    \mathcal{G}_{k,k}+\rho\sum\limits_{k'\neq k}p_{k'}\mathcal{G}_{k,k'}+\sigma_w^2},
\end{align}
where $\mathcal{G}_{k1,k2}=\|\mathbf{u}_{k_1}^{\mathrm{H}}\sqrt{\mathbf{\Upsilon}}
\mathbf{H}\mathbf{F}\mathbf{\Phi}\mathbf{v}_{k_2}\|^2$. Thus, the spectral efficiency, denoted as $\mathcal{S}$, is given by
\begin{align}\label{Hybrid_Beamforming_Design_2_1}
    \mathcal{S}=\sum\limits_{k=1}^K\log_2\left(1+\gamma_k\right),
\end{align}
and the energy efficiency, denoted as $\mathcal{E}$, is given by
\begin{align}\label{Hybrid_Beamforming_Design_2_2}
    \mathcal{E}=\frac{B\mathcal{S}}{\mathcal{P}}
    =\frac{B\sum_{k=1}^K\log_2\left(1+\gamma_k\right)}
    {\frac{1}{\zeta}\rho+P_{\mathrm{Syn}}+N_{\mathrm{RF}}P_{\mathrm{RF}}+
    N_{\mathrm{PS}}P_{\mathrm{PS}}+N_{\mathrm{SW}}P_{\mathrm{SW}}},
\end{align}
where $B$ is the system's bandwidth, $\mathcal{P}=\frac{1}{\zeta}\rho+P_{\mathrm{Syn}}+N_{\mathrm{RF}}P_{\mathrm{RF}}+
N_{\mathrm{PS}}P_{\mathrm{PS}}+N_{\mathrm{SW}}P_{\mathrm{SW}}$ is total power required for achieving the spectral efficiency $\mathcal{S}$ with $\zeta$ being the efficiency of the power amplifiers, $P_\mathrm{Syn}$ being the power consumption of the frequency synthesizer, $N_\mathrm{RF}$, $N_\mathrm{PS}$ and $N_\mathrm{SW}$ being the number of RF chains, phase shifters and switches, respectively. Finally, $P_\mathrm{RF}$, $P_\mathrm{PS}$ and $P_\mathrm{SW}$ are the power consumption of each RF chain, phase shifter and switch, respectively.

Our objective is to maximize the energy efficiency by optimizing the total transmit power $\rho$, the power sharing ratio $\mathbf{P}$, the digital beamformer $\mathbf{V}$ and $\mathbf{U}$ at the transmitter and the receiver, respectively, as well as the holographic beamformer $\mathbf{F}$. Thus, the optimization problem in the switch-controlled holographic beamforming architecture can be formulated as
\begin{align}
    \text{(P1)}\quad &\max_{\rho,\mathbf{P},\mathbf{V},\mathbf{U},\mathbf{F}}\ \frac{B\sum_{k=1}^K\log_2\left(1+\gamma_k\right)}{\mathcal{P}}\\
    \text{s.t.}&\quad \rho\leq P_{\mathrm{max}},\\
    &\quad \sum_{k=1}^{K}p_k=1,\\
    &\quad \left\|\mathbf{v}_k\|^2=1,\|\mathbf{u}_k\right\|^2=1,\ k=1,2,\cdots,K,\\
    &\quad \xi_n\in\left\{0,1\right\},\ n=1,2,\cdots,N.
\end{align}
respectively, where $P_{\mathrm{max}}$ is the maximum transmit power budget. To solve problem (P1) efficiently, we decompose them into three sub-problems as follows.

\subsection{Holographic Beamformer Design}\label{RHS_based_Beamforming}
In the holographic beamformer, we aim for maximizing the sum of the eigen-channel gains of all UEs based on sum-path-gain maximization (SPGM) criterion \cite{ning2020beamforming}. We denote the equivalent channel at the baseband corresponding to UE-$k$ as $\mathbf{g}_k^\mathrm{H}=\mathbf{h}_k^{\mathrm{H}}\mathbf{F}\mathbf{\Phi}$. Thus, the optimization problem can be formulated as
\begin{align}\label{Hybrid_Beamforming_Design_14_b}
    \text{(P2)}\quad &\max_{\mathbf{F}}\ \sum_{k=1}^{K}\left\|\mathbf{g}_k\right\|_2^2\\
    \text{s.t.}&\quad \xi_n\in\left\{0,1\right\},\ n=1,2,\cdots,N,
\end{align}
for the switch-controlled holographic beamformer. In (\ref{Hybrid_Beamforming_Design_14_b}), we have
\begin{align}
    \notag\sum_{k=1}^{K}\left\|\mathbf{g}_k\right\|_2^2
    \notag=&\sum_{k=1}^{K}\left\|\mathbf{h}_k^{\mathrm{H}}\mathbf{F}\mathbf{\Phi}\right\|_2^2\\
    =&\boldsymbol{\xi}^\mathrm{T}\left(\sum_{k=1}^K\sum_{m=1}^M
    \frac{1}{N}\mathbf{q}_{k,m}\mathbf{q}_{k,m}^{\mathrm{H}}\right)\boldsymbol{\xi},
\end{align}
where $\mathbf{q}_{k,m}=[\mathrm{e}^{-j\phi_{1,m}}h_{k,1},\mathrm{e}^{-j\phi_{2,m}}h_{k,2},
\cdots,\mathrm{e}^{-j\phi_{N,m}}h_{k,N}]^\mathrm{H}$ and $\boldsymbol{\xi}=[\xi_1,\xi_2,\cdots\xi_N]^{\mathrm{T}}$. Upon introducing the matrix $\mathbf{Q}=\sum_{k=1}^K\sum_{m=1}^M\frac{1}{N}\mathbf{q}_{k,m}\mathbf{q}_{k,m}^\mathrm{H}$, the optimization problem (P2) can be written as
\begin{align}
    \text{(P3)}\quad &\max_{\boldsymbol{\xi}}\ \boldsymbol{\xi}^\mathrm{T}\mathbf{Q}\boldsymbol{\xi}\\
    \text{s.t.}&\quad \boldsymbol{\xi}\in\mathbb{R}^{N\times1},\ \xi_n\in\left\{0,1\right\},\ n=1,2,\cdots,N.
\end{align}

Since $\mathbf{Q}\succeq\mathbf{0}$, the problem (P3) is a quadratic programming and the optimal solution is located at the vertex. Therefore, the optimization problem associated with the constraint $0\leq\xi_n\leq1$ and that with the constraint $\xi_n\in\{0,1\}$ have the same optimal solution, which means that the switch-controlled RHS beamforming architecture leads to the same performance as the continuous amplitude-controlled RHS beamforming architecture. Due to the module-one constraint, (P3) is a non-convex optimization problem. Although (P3) can be solved by the semidefinite relaxation (SDR) method, it has high calculation complexity \cite{chi2017convex}. Therefore, we employ the low-complexity ED method as follows.

Upon defining $\boldsymbol{\xi}=\frac{1}{2}(\widetilde{\boldsymbol{\xi}}+\mathbf{1}_N)$ with $\widetilde{\xi}_n\in\{-1,1\}$ in the problem (P3), we can get
\begin{align}
    \boldsymbol{\xi}^\mathrm{T}\mathbf{Q}\boldsymbol{\xi}
    =\frac{1}{4}\left(\widetilde{\boldsymbol{\xi}}^\mathrm{T}\mathbf{Q}\widetilde{\boldsymbol{\xi}}
    +\mathbf{1}_N^\mathrm{T}\mathbf{Q}\widetilde{\boldsymbol{\xi}}
    +\widetilde{\boldsymbol{\xi}}^\mathrm{T}\mathbf{Q}\mathbf{1}_N
    +\mathbf{1}_N^\mathrm{T}\mathbf{Q}\mathbf{1}_N\right).
\end{align}
Since the value of $\mathbf{1}_N^\mathrm{T}\mathbf{Q}\mathbf{1}_N$ is independent of the variable $\widetilde{\boldsymbol{\xi}}$, the problem of maximizing $\boldsymbol{\xi}^\mathrm{T}\mathbf{Q}\boldsymbol{\xi}$ is equivalent to that of maximizing $\widetilde{\boldsymbol{\xi}}^\mathrm{T}\mathbf{Q}\widetilde{\boldsymbol{\xi}}
+\mathbf{1}_N^\mathrm{T}\mathbf{Q}\widetilde{\boldsymbol{\xi}}
+\widetilde{\boldsymbol{\xi}}^\mathrm{T}\mathbf{Q}\mathbf{1}_N$. Upon introducing the auxiliary variable $a\in\{-1,1\}$, and defining $\widetilde{\boldsymbol{\xi}}=a\mathbf{z}$ and $\widetilde{\mathbf{z}}=[\mathbf{z}^\mathrm{T},a]^\mathrm{T}$, the problem (P3) can be presented as
\begin{align}
    \text{(P4)}\quad &\max_{\widetilde{\mathbf{z}}}\
    \widetilde{\mathbf{z}}^\mathrm{T}\widetilde{\mathbf{Q}}\widetilde{\mathbf{z}}\\
    \text{s.t.}&\quad \widetilde{\mathbf{z}}\in\mathbb{R}^{N\times1},\ \widetilde{z}_n\in\left\{-1,1\right\},\ n=1,2,\cdots,N+1,
\end{align}
where we have:
\begin{align}
    \widetilde{\mathbf{Q}}=\left[\begin{array}{cc}
     \mathbf{Q} & \mathbf{Q}\mathbf{1}_N \\
     \mathbf{1}_N^\mathrm{T}\mathbf{Q} & 0
    \end{array}\right].
\end{align}
Due to the module-one constraint, (P4) is a non-convex optimization problem and ED method for solving (P4) is presented as follows.

According to the constraint of (P4), we have $\|\widetilde{\mathbf{z}}\|^2=N+1$. We denote the eigen-decomposition of $\widetilde{\mathbf{Q}}$ as $\widetilde{\mathbf{Q}}=\sum_{n=1}^{N+1}\widetilde{\lambda}_n\widetilde{\boldsymbol{\nu}}_n
\widetilde{\boldsymbol{\nu}}_n^{\mathrm{H}}$, where $\widetilde{\lambda}_1,\widetilde{\lambda}_2,\cdots,\widetilde{\lambda}_{N+1}$ are the eigen-values of $\widetilde{\mathbf{Q}}$ in a descending order, and $\widetilde{\boldsymbol{\nu}}_1,\widetilde{\boldsymbol{\nu}}_2,\cdots,
\widetilde{\boldsymbol{\nu}}_{N+1}$ are the corresponding eigen-vectors. Based on the Rayleigh quotient, defined in \cite{chi2017convex}, the optimal solution of $f(\mathbf{x})=\frac{\mathbf{x}^{\mathrm{H}}\widetilde{\mathbf{Q}}\mathbf{x}}
{\mathbf{x}^{\mathrm{H}}\mathbf{x}}$ is $\mathbf{x}^*=c\widetilde{\boldsymbol{\nu}}_1$, where $c$ is an arbitrary non-zero constant and $\widetilde{\boldsymbol{\nu}}_1$ is the eigen-vector corresponding to the largest eigen-value of $\widetilde{\mathbf{Q}}$. Thus, based on the relaxation and projection method of \cite{zhang2020sum,zhou2021stochastic}, the optimal solution of (P4), denoted as $\widetilde{\mathbf{z}}^*$, is $\widetilde{\mathbf{z}}^*=\mathrm{e}^{j\widetilde{\boldsymbol{\nu}}_1}$.

In the switch-controlled holographic beamforming architecture, we have $N_\mathrm{RF}=M$, $N_\mathrm{PS}=0$, $N_\mathrm{SW}=N$. Then the total power consumption is $\mathcal{P}=\frac{1}{\zeta}\rho+P_{\mathrm{Syn}}+MP_\mathrm{RF}+NP_\mathrm{SW}$.

\subsection{Digital Beamformer Design}\label{Digital_Beamforming}
Given the holographic beamformer $\mathbf{F}$, we can get the baseband equivalent channel as
\begin{align}
    \mathbf{G}=\sqrt{\mathbf{\Upsilon}}\mathbf{H}\mathbf{F}\mathbf{\Phi}.
\end{align}
Then the optimization problem can be formulated as
\begin{align}\label{Hybrid_Beamforming_Design_5}
    \notag\text{(P5)}\quad &\max_{\rho,\mathbf{P},\mathbf{V},\mathbf{U}}\ \sum_{k=1}^K\log_2\left(1+\right.\\
    &\left.\ \frac{\rho\varepsilon_\mathrm{r}\varepsilon_\mathrm{t}p_k\mathcal{G}_{k,k}}
    {\rho\left(1-\varepsilon_\mathrm{r}\varepsilon_\mathrm{t}\right)p_k
    \mathcal{G}_{k,k}+\rho\sum_{k'\neq k}p_{k'}
    \mathcal{G}_{k,k'}+\sigma_w^2}\right)\\
    \text{s.t.}&\quad \sum_{k=1}^{K}p_k=1,\\
    &\quad \left\|\mathbf{v}_k\right\|^2=1,\left\|\mathbf{u}_k\right\|^2=1,\ k=1,2,\cdots,K.
\end{align}

To eliminate the inter-user interference, the singular value decomposition (SVD) can be employed for the digital beamformer as follows\footnote{Although it is more practical to employ the zero-forcing (ZF) method for multi-user systems, the SVD method can help us unveil the performance limit of the RHS-based multi-user systems.}. We denote the SVD of the baseband equivalent channel of $\mathbf{G}$ as
\begin{align}\label{Hybrid_Beamforming_Design_6}
    \mathbf{G}=\mathbf{U}_{\mathbf{G}}\mathbf{\Lambda}\mathbf{V}_{\mathbf{G}}^{\mathrm{H}},
\end{align}
where $\mathbf{\Lambda}=\{\lambda_1,\lambda_2,\cdots,\lambda_K\}$ is a diagonal matrix associated with diagonal elements that are the singular values in descending order, while $\mathbf{U}_{\mathbf{G}}\in\mathbb{C}^{K\times K}$ and $\mathbf{V}_{\mathbf{G}}\in\mathbb{C}^{N\times K}$ are complex unitary matrices. Thus, the digital precoding matrix $\mathbf{V}$ of the transmitter and the combining matrix $\mathbf{U}^{\mathrm{H}}$ of the receiver are designed as $\mathbf{V}=\mathbf{V}_{\mathbf{G}}$ and $\mathbf{U}^{\mathrm{H}}=\mathbf{U}_{\mathbf{G}}^{\mathrm{H}}$, respectively.

\subsection{Power Sharing Ratio and Total Transmit Power Optimization}\label{Power_Allocation_Ratio}
According to (\ref{Hybrid_Beamforming_Design_5}) and (\ref{Hybrid_Beamforming_Design_6}), it may be readily shown that $\|\mathbf{u}_k^{\mathrm{H}}\mathbf{G}\mathbf{v}_k\|^2=\upsilon_k\lambda_k^2$ and $\|\mathbf{u}_k^{\mathrm{H}}\mathbf{G}\mathbf{v}_{k'}\|^2=0$ if $k\neq k'$. Then (P1) can be expressed as
\begin{align}
    \text{(P6)}\quad &\max_{\rho,\mathbf{P}}\ \frac{B\sum_{k=1}^K\log_2\left(1+
    \frac{\rho\lambda_k^2\varepsilon_\mathrm{r}\varepsilon_\mathrm{t}p_k}
    {\rho\lambda_k^2\left(1-\varepsilon_\mathrm{r}\varepsilon_\mathrm{t}\right)p_k
    +\sigma_w^2}\right)}
    {\frac{1}{\zeta}\rho+P_{\mathrm{Syn}}+MP_{\mathrm{RF}}+NP_{\mathrm{SW}}}\\
    \text{s.t.}&\quad \rho\leq P_{\mathrm{max}},\\
    &\quad \sum_{k=1}^{K}p_k=1.
\end{align}
Since (P6) is a non-convex problem, we can decouple them into a pair of sub-problems and optimize them iteratively. Specifically, when the total transmit power $\rho$ is given, the problems (P6) aims for optimizing the power sharing ratio $\mathbf{P}$, represented as
\begin{align}
    \text{(P6.a)}\quad &\max_{\mathbf{P}}\ \sum\limits_{k=1}^K\log_2\left(1+
    \frac{\rho\lambda_k^2\varepsilon_\mathrm{r}\varepsilon_\mathrm{t}p_k}
    {\rho\lambda_k^2\left(1-\varepsilon_\mathrm{r}\varepsilon_\mathrm{t}\right)p_k
    +\sigma_w^2}\right)\\
    \text{s.t.}&\quad \sum_{k=1}^{K}p_k=1.
\end{align}
By contrast, when the power sharing ratio $\mathbf{P}$ is given, the problem (P6) aims for optimizing the total transmit power $\rho$, represented as
\begin{align}\label{Hybrid_Beamforming_Design_10_b}
    \text{(P6.b)}\quad &\max_{\rho}\ \frac{B\sum_{k=1}^K\log_2\left(1+
    \frac{\rho\lambda_k^2\varepsilon_\mathrm{r}\varepsilon_\mathrm{t}p_k}
    {\rho\lambda_k^2\left(1-\varepsilon_\mathrm{r}\varepsilon_\mathrm{t}\right)p_k
    +\sigma_w^2}\right)}
    {\frac{1}{\zeta}\rho+P_{\mathrm{Syn}}+MP_{\mathrm{RF}}+NP_{\mathrm{SW}}}\\
    \text{s.t.}&\quad \rho\leq P_{\mathrm{max}}.
\end{align}
By iteratively optimizing the power sharing ratio $\mathbf{P}$ and the total transmit power $\rho$ until convergence, a nearly-optimal solution can be found. The details of the power sharing ratio optimization and the total transmit power optimization algorithm are given as follows.

\subsubsection{Power sharing ratio optimization}
When the hardware is ideal, i.e. $\varepsilon_\mathrm{t}=1$ and $\varepsilon_\mathrm{r}=1$, the problem (P6.a) can be solved by the state-of-the-art water-filling method~\cite{tse2005fundamentals}. By contrast, in the case of realistic non-ideal hardware, i.e. $\varepsilon_\mathrm{t}<1$ or $\varepsilon_\mathrm{r}<1$, the problem (P6.a) can be solved by using the Lagrange multiplier method of the calculus of variations~\cite{hampton2013introduction}. Hence, we can formulate our problem as
\begin{align}
    J(\mathbf{P})=\sum_{k=1}^{K}\log_2
    \left(1+\frac{\rho\lambda_k^2\varepsilon_\mathrm{r}\varepsilon_\mathrm{t}p_k}
    {\rho\lambda_k^2\left(1-\varepsilon_\mathrm{r}\varepsilon_\mathrm{t}\right)p_k
    +\sigma_w^2}\right)+{b}\sum_{k=1}^{K}p_k,
\end{align}
where ${b}$ is the Lagrange multiplier. Then, we take the partial derivative of $J(\mathbf{P})$ with respect to the power sharing variables to be optimized $p_1,p_2,\cdots,p_K$ and set them equal to zero, which results in
\begin{align}\label{Hybrid_Beamforming_Design_11}
    \notag\frac{J(\mathbf{P})}{\partial{p_k}}=
    &\frac{\rho\lambda_k^2\varepsilon_\mathrm{r}\varepsilon_\mathrm{t}\sigma_w^2}
    {\ln2\cdot\left(\rho\lambda_k^2p_k+\sigma_w^2\right)
    \left(\rho\lambda_k^2\left(1-\varepsilon_\mathrm{r}\varepsilon_\mathrm{t}\right)p_k
    +\sigma_w^2\right)}+{b}\\
    =&0.
\end{align}
According to (\ref{Hybrid_Beamforming_Design_11}), $p_k$ can be expressed as
\begin{align}
    p_k=\max\big\{f_k(b),0\big\},
\end{align}
where $f_k(b)=\frac{\sqrt{\varepsilon_\mathrm{r}^2\varepsilon_\mathrm{t}^2
-\frac{\frac{4\rho}{\sigma_w^2}\lambda_k^2(1-\varepsilon_\mathrm{r}\varepsilon_\mathrm{t})
\varepsilon_\mathrm{r}\varepsilon_\mathrm{t}}{\ln2\cdot{b}}}
-(2-\varepsilon_\mathrm{r}\varepsilon_\mathrm{t})}
{\frac{2\rho}{\sigma_w^2}\lambda_k^2(1-\varepsilon_\mathrm{r}\varepsilon_\mathrm{t})}$ and $\max\{x,0\}$ represents the maximum value between $x$ and 0. Furthermore, since $\sum_{k=1}^{K}p_k=1$, we have
\begin{align}
    \sum_{k=1}^{K}\max\left\{f_k(b),0\right\}=1.
\end{align}
Algorithm~\ref{algorithm_1} presents the details of deriving the power sharing ratio of $p_1,p_2,\cdots,p_K$, where the value of ${b}$ in the equation $\sum\limits_{k=1}^{K-\tau}
f_k(b)=1$ can be numerically calculated by the bisection method.

\begin{algorithm}[!t]
\caption{Power allocation method conceived for maximizing the energy efficiency.}\label{algorithm_1}
\begin{algorithmic}[1]
\REQUIRE
    Hardware quality factors $\varepsilon_\mathrm{t}$ and $\varepsilon_\mathrm{r}$, transmit power $\rho$, baseband channel gain $\lambda_1,\lambda_2,\cdots,\lambda_K$, and additive noise power $\sigma_w^2$.
    \STATE
        Initial $\tau=0$.
    \STATE
        \textbf{repeat}
    \STATE
        \quad Numerically calculate $b$ in the equation $\sum\limits_{k=1}^{K-\tau}f_k(b)=1$.
    \STATE
        \quad Update $p_k=f_k(b)$ for $k=1,2,\cdots,K-\tau$.
    \STATE
        \quad Update $p_k=0$ for $k=K-\tau+1,K-\tau+2\cdots,K$.
    \STATE
        \quad $\tau\leftarrow\tau+1$.
    \STATE
        \textbf{until} $p_k\geq0$ for $k=1,2,\cdots,K$.
\ENSURE
    The optimal power allocation ratio, denoted as $\mathbf{P}^*=\mathbf{Diag}\{p_1^*,p_2^*,\cdots,p_K^*\}$.
\end{algorithmic}
\end{algorithm}

\subsubsection{Total transmit power optimization}
In (\ref{Hybrid_Beamforming_Design_10_b}), we denote the energy efficiency of the switch-controlled holographic beamformer as $\mathcal{E}$. Since the functions of $\mathcal{E}$ with respect to the transmit power $\rho$ are concave, the optimal transmit power $\rho$ can be found by the gradient descent method, which is given in Algorithm~\ref{algorithm_2}, where the gradient of $\mathcal{E}$ with respect to the transmit power $\rho$ is formulated as:
\begin{align}
    \notag\frac{\partial\mathcal{E}}{\partial\rho}
    =&\frac{\partial\frac{B\sum_{k=1}^K\log_2\left(1+
    \frac{\rho\lambda_k^2\varepsilon_\mathrm{r}\varepsilon_\mathrm{t}p_k}
    {\rho\lambda_k^2\left(1-\varepsilon_\mathrm{r}\varepsilon_\mathrm{t}\right)
    p_k+\sigma_w^2}\right)}
    {\frac{1}{\zeta}\rho+P_{\mathrm{Syn}}+MP_{\mathrm{RF}}+NP_{\mathrm{SW}}}}{\partial\rho}\\
    \notag=&\frac{B}
    {\ln2\cdot\left(\frac{1}{\zeta}\rho+P_{\mathrm{Syn}}+MP_{\mathrm{RF}}
    +NP_{\mathrm{SW}}\right)^2}\cdot\\
    \notag&\left(\sum_{k=1}^K\frac{\left(\frac{1}{\zeta}\rho+P_{\mathrm{Syn}}
    +MP_{\mathrm{RF}}+NP_{\mathrm{SW}}\right)
    \lambda_k^2\varepsilon_\mathrm{r}\varepsilon_\mathrm{t}p_k\sigma_w^2}
    {\left(\rho\lambda_k^2p_k+\sigma_w^2\right)
    \left(\rho\lambda_k^2\left(1-\varepsilon_\mathrm{r}\varepsilon_\mathrm{t}\right)p_k
    +\sigma_w^2\right)}\right.\\
    &\left.-\frac{1}{\zeta}\sum_{k=1}^K\ln\left(1+
    \frac{\rho\lambda_k^2\varepsilon_\mathrm{r}\varepsilon_\mathrm{t}p_k}
    {\rho\lambda_k^2\left(1-\varepsilon_\mathrm{r}\varepsilon_\mathrm{t}\right)p_k
    +\sigma_w^2}\right)\right).
\end{align}

\begin{algorithm}[!t]
\caption{Gradient descent method conceived for optimizing the total transmit power.}\label{algorithm_2}
\begin{algorithmic}[1]
\REQUIRE
    Hardware quality factors $\varepsilon_\mathrm{t}$ and $\varepsilon_\mathrm{r}$, power allocation ratio $p_1,p_2,\cdots,p_K$, baseband channel gain $\lambda_1,\lambda_2,\cdots,\lambda_K$, additive noise power $\sigma_w^2$, iteration step $\iota$, precision error $\epsilon$ and maximum transmit power budget $P_\mathrm{max}$.
    \STATE
        Initial $\tau=0$, and transmit power $\rho^{(\tau)}=P_\mathrm{max}$.
    \STATE
        \textbf{if} $\frac{\partial\mathcal{E}}{\partial\rho^{(\tau)}}\geq0$
    \STATE
        \quad The optimal transmit power $\rho^*=P_\mathrm{max}$.
    \STATE
        \textbf{else}
    \STATE
        \textbf{repeat}
    \STATE
        \quad $\rho^{(\tau+1)}=\rho^{(\tau)}+\iota\cdot
        \frac{\partial\mathcal{E}}{\partial\rho^{(\tau)}}$.
    \STATE
        \quad $\tau\leftarrow\tau+1$.
    \STATE
        \textbf{until} $\left|\frac{\partial\mathcal{E}}{\partial\rho^{(\tau)}}\right|<\epsilon$.
\ENSURE
    The optimal transmit power $\rho^*$.
\end{algorithmic}
\end{algorithm}

\subsection{Theoretical Analysis}
Firstly, the impact of hardware impairments on the spectral efficiency and energy efficiency is analyzed. Then, we present the time complexity of our proposed hybrid beamformer design.

\subsubsection{Impact of Hardware Impairments}
We investigate the impact of the hardware impairments on the transceiver's spectral efficiency and energy efficiency. According to (\ref{Hybrid_Beamforming_Design_1}) and (\ref{Hybrid_Beamforming_Design_2_1}), we can get that in the high signal-to-noise ratio (SNR) region, the spectral efficiency is limited by the hardware impairments at the transceivers, which results in a performance floor. This is due to the fact that increasing the transmit power increases both the desired signal and the signal distortion resulting from the HWI. Specifically, when the hardware of the transceiver is imperfect, the spectral efficiency attained in the high-SNR region tends to saturation at:
\begin{align}\label{Theoretical_Analysis_1}
    \mathcal{S}=K\log_2\left(1+\frac{\varepsilon_\mathrm{r}\varepsilon_\mathrm{t}}
    {1-\varepsilon_\mathrm{r}\varepsilon_\mathrm{t}}\right),
\end{align}
which can be obtained by setting $\rho\rightarrow\infty$ in (\ref{Hybrid_Beamforming_Design_1}). Furthermore, according to (\ref{Hybrid_Beamforming_Design_2_2}) and (\ref{Theoretical_Analysis_1}), we can formulate the energy efficiency in the high-SNR region as
\begin{align}
    \mathcal{E}=\frac{BK\log_2\left(1+\frac{\varepsilon_\mathrm{r}\varepsilon_\mathrm{t}}
    {1-\varepsilon_\mathrm{r}\varepsilon_\mathrm{t}}\right)}
    {\frac{1}{\zeta}\rho+P_{\mathrm{Syn}}+N_{\mathrm{RF}}P_{\mathrm{RF}}+
    N_{\mathrm{PS}}P_{\mathrm{PS}}+N_{\mathrm{SW}}P_{\mathrm{SW}}}.
\end{align}
By optimizing the transmit power $\rho$ to maximize the energy efficiency, we can formulate the energy efficiency upper bound of the switch-controlled holographic beamforming architecture, which is given by
\begin{align}\label{Theoretical_Analysis_4}
    \overline{\mathcal{E}}
    =\frac{BK\log_2\left(1+\frac{\varepsilon_\mathrm{r}\varepsilon_\mathrm{t}}
    {1-\varepsilon_\mathrm{r}\varepsilon_\mathrm{t}}\right)}
    {P_{\mathrm{Syn}}+MP_{\mathrm{RF}}+NP_{\mathrm{SW}}}.
\end{align}

\subsubsection{Computational complexity}
The computational complexity of the proposed hybrid beamformer design
depends on the number of iterations in the alternating maximization, denoted as $T_a$, and on the computational complexity required to solve each sub-problem. As for the sub-problem of holographic beamformer design, it can be seen that the optimization of the RHS coefficients depends on the number of RHS elements $N$, which can be seen to scale with $N^3$ \cite{chi2017convex}. As for the sub-problem of digital beamformer design, each SVD process in (\ref{Hybrid_Beamforming_Design_6}) requires $NK^2$ floating-point operations \cite{li2019tutorial}. Moreover, the sub-problem of the power sharing ratio optimization in Algorithm~\ref{algorithm_1} updates the power ratio factors at most $K$ times, and in each update the equation $\sum\limits_{k=1}^{K-\tau}f_k(b)=1$ is numerically calculated \cite{ben2001lectures}. Since it only depends on the single variable of $b$, the sub-problem of the power sharing ratio optimization in Algorithm~\ref{algorithm_1} has the complexity order of $\mathcal{O}(K)$. Since the sub-problem of the total transmit power optimization in Algorithm~\ref{algorithm_2} iterates $\tau$ times and each iteration solves a single-variable gradient update, the sub-problem of the total transmit power optimization in Algorithm~\ref{algorithm_2} has the complexity order of $\mathcal{O}(\tau)$. Since $N>K$, the overall computational complexity of our proposed hybrid beamformer design is $\mathcal{O}(T_a(N^3+\tau))$.

\section{Numerical and Simulation Results}\label{Numerical_and_Simulation_Results}
In this section, the simulation results characterizing the energy efficiency of the switch-controlled RHS beamformers is presented. Similarly to~\cite{ni2015hybrid}, we denote the average large-scale fading of all users as $\overline{\upsilon}$, and the large-scale fading of the $k$th user as $\upsilon_k=\beta_k\overline{\upsilon}$, where the factors $\beta_1,\beta_2,\cdots,\beta_K$ are uniformly distributed in [0.5, 1.5]. Unless otherwise specified, the simulation parameters are set as follows: the bandwidth $B=20\mathrm{MHz}$, the number of users is $K=8$, the number of RF chains is $M=8$, the number of RHS elements is $N=16\times16$, $\overline{\upsilon}=-80\mathrm{dB}$, $\sigma_w^2=-90\mathrm{dBm}$, and the iterative times of the alternating optimization is $\tau=4$. Similarly to \cite{ni2015hybrid}, the mmWave propagation channel parameters are set as: the number of clusters is $L_{C}=8$, the number of paths in each cluster is $L_{P}=10$. The mean values of the elevation angles $\mu_{\psi_{c}^k}$ are randomly distributed in $[0^\circ,180^\circ]$, the mean values of the azimuth angles $\mu_{\varphi_{c}^k}$ are randomly distributed in $[0^\circ,360^\circ]$, the elevation angle spreads are $\sigma_{\psi_{c,p}^k}=7.5^\circ$, and the azimuth angle spreads are $\sigma_{\varphi_{c,p}^k}=7.5^\circ$, for $c=1,2,\cdots,L_{C}$, $p=1,2,\cdots,L_{P}$ and $k=1,2,\cdots,K$. According to~\cite{payami2018phase}, \cite{garcia2016hybrid}, the power consumption parameters are set as: the efficiency of the power amplifiers is $\zeta=0.39$, the power consumption of the frequency synthesizer is $P_\mathrm{Syn}=2\mathrm{W}$, while the power consumption of the RF chains, phase shifters
and switches are $P_\mathrm{RF}=1\mathrm{W}$, $P_\mathrm{PS}=30\mathrm{mW}$ and $P_\mathrm{SW}=1\mathrm{mW}$, respectively.

For comparison, we employ the following state-of-the-art beamforming schemes as benchmarks:
\begin{enumerate}
  \item \textit{Fully digital beamformer:} The signals are processed by a pure digital beamformer having $N$ RF chains. Thus, $N$ RF chains dissipate power. Thus, the total power consumption in the fully digital beamformer is $\mathcal{P}=\frac{1}{\zeta}\rho+P_{\mathrm{Syn}}+NP_\mathrm{RF}$.
  \item \textit{Fully-connected phase shift array (PSA) beamformer:} The signals are processed by a digital beamformer having $M$ RF chains, followed by an analogue beamformer equipped with $N$ antennas, where the $M$ RF chains and the $N$ antennas are fully-connected by an $N\times M$ PSA. The analogue beamformer is designed based on the right singular matrix of the channel~\cite{payami2016hybrid}, \cite{payami2018hybrid}, \cite{payami2018phase}. Specifically, we denote the right singular matrix as $\mathbf{V}\in\mathbb{C}^{N\times M}$. The coefficient of the phase shifter connecting the $m$th RF chain and the $n$th transmit antenna, denoted as $f^{\mathrm{(FPSA)}}_{n,m}$, is designed as $f^{\mathrm{(FPAS)}}_{n,m}=\frac{1}{\sqrt{N}}\mathrm{e}^{j\angle V_{n,m}}$. Thus, a total of $M$ RF chains and $MN$ phase shifters consume power during signal processing. Thus, the total power consumption in the fully-connected PSA beamformer is $\mathcal{P}=\frac{1}{\zeta}\rho+P_{\mathrm{Syn}}+MP_\mathrm{RF}+MNP_\mathrm{PS}$.
  \item \textit{Sub-connected PSA beamformer:} The signals are processed by a digital beamformer having $M$ RF chains, followed by an analogue beamformer associated with $N$ antennas, where the $M$ RF chains and the $N$ antennas are `sub-connected', i.e. each RF chain is connected to $\frac{N}{M}$ antennas by an $\frac{N}{M}\times1$ PSA. The analogue beamformer is designed based on the right singular matrix of the channel~\cite{payami2016hybrid}, \cite{payami2018hybrid}. The coefficient of the phase shifter connecting the $m$th RF chain and the $n$th transmit antenna, denoted as $f^{\mathrm{(FPSA)}}_{n,m}$, is designed as $f^{\mathrm{(SPAS)}}_{n,m}=\frac{1}{\sqrt{N}}\mathrm{e}^{j\angle V_{n,m}}$ if the $m$th RF chain and the $n$th transmit antenna are connected, and $f^{\mathrm{(SPAS)}}_{n,m}=0$ otherwise. Thus, a total of $M$ RF chains and $N$ phase shifters consume power. Thus, the total power consumption in the sub-connected PSA beamformer is $\mathcal{P}=\frac{1}{\zeta}\rho+P_{\mathrm{Syn}}+MP_\mathrm{RF}+NP_\mathrm{PS}$.
  \item \textit{PSA with switches beamformer:} To further reduce the number of phase shifters activated, the method of PAS with switches is proposed in~\cite{payami2016hybrid}, \cite{payami2018hybrid}. Based on the sub-connected PSA circuit, each phase shifter is connected to a switch. Some phase shifters associated with low-magnitude element in the right singular matrix are deactivated by turning off the corresponding switch. Specifically, the state of the switch connecting the $m$th RF chain and the $n$th transmit antenna, denoted as $f^{\mathrm{(S)}}_{n,m}$, is defined as
    \begin{align}
            f^{\mathrm{(S)}}_{n,m}=\left\{\begin{array}{ll}
            1, & \text{if }V_{n,m}> a_{(\kappa)} \\
            0, & \text{if }V_{n,m}\leq a_{(\kappa)}
        \end{array}\right.,
    \end{align}
where $f^{\mathrm{(S)}}_{n,m}=1$ represents that the switch is turned on, while $f^{\mathrm{(S)}}_{n,m}=1$ indicates that the switch is turned off. Furthermore, $a_{(\kappa)}$ is the threshold, which ensures that $(1-\kappa)N$ phase shifters are activated while the remaining $\kappa N$ phase shifters are deactivated. Thus, a total of $M$ RF chains, $(1-\kappa)N$ phase shifters and $N$ switches consume power. Thus, the total power consumption in the PSA with switches beamformer is $\frac{1}{\zeta}\rho+P_{\mathrm{Syn}}
+MP_{\mathrm{RF}}+(1-\kappa)NP_{\mathrm{PS}}+NP_{\mathrm{SW}}$.
\end{enumerate}

Explicitly, the fully digital and fully-connected PSA schemes have the highest power consumption. The power consumption is reduced by the sub-connected PSA beamformer. Furthermore, the PSA with switches scheme and the switch-controlled RHS beamforming scheme have very low power consumption due to the employment of low-power switches allowing us to reduce the number of phase shifters.

\begin{figure}[!t]
    \centering
    \subfloat[Hardware quality factor $\varepsilon_\mathrm{t}=\varepsilon_\mathrm{r}=1$.]{\begin{minipage}{1\linewidth}
        \centering
        \includegraphics[width=2.8in]{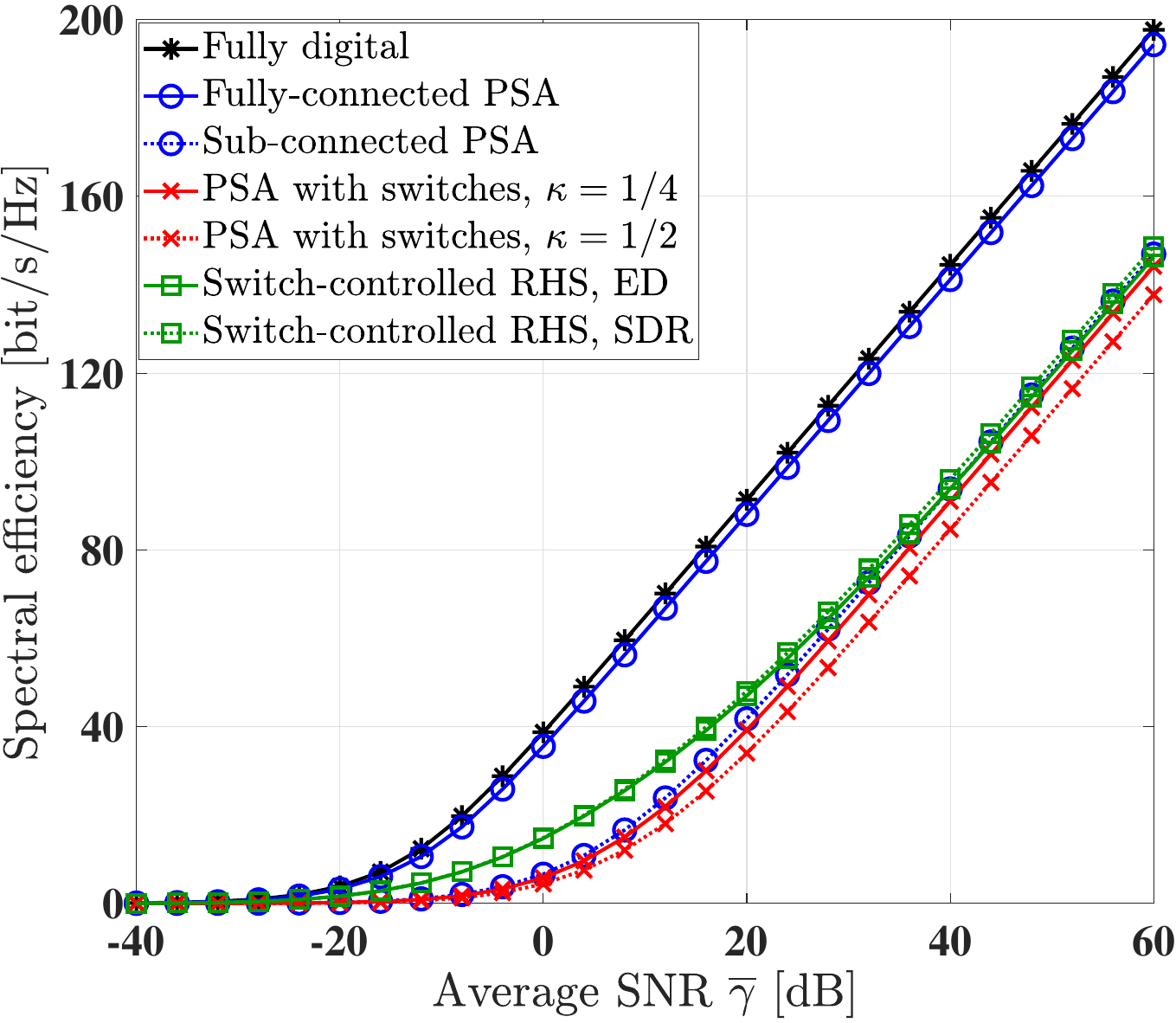}
    \end{minipage}}\\
    \subfloat[Hardware quality factors $\varepsilon_\mathrm{t}=\varepsilon_\mathrm{r}=0.8$.]{\begin{minipage}{1\linewidth}
        \centering
        \includegraphics[width=2.8in]{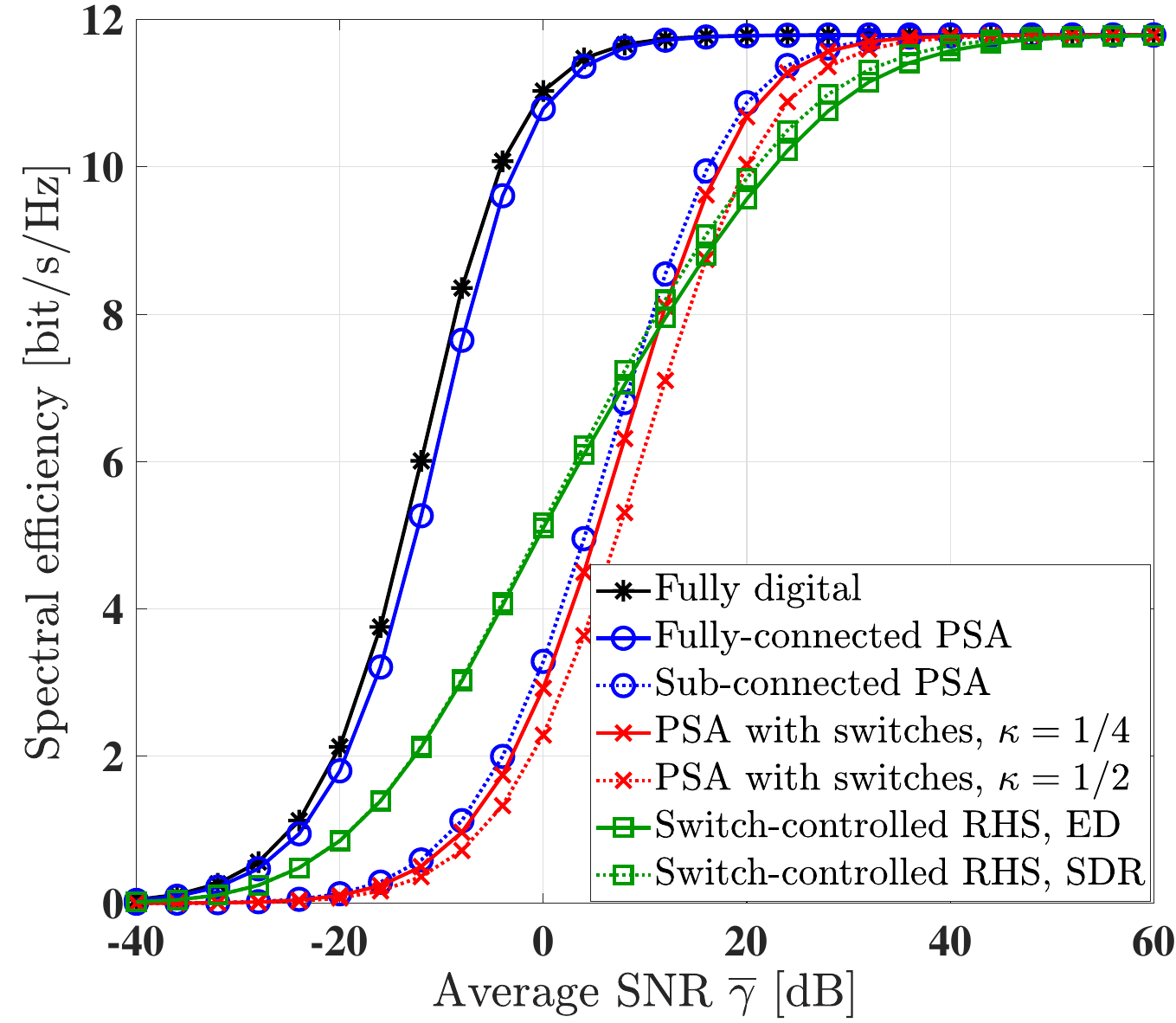}
    \end{minipage}}\\
    \subfloat[Hardware quality factors $\varepsilon_\mathrm{t}=\varepsilon_\mathrm{r}=0.6$.]{\begin{minipage}{1\linewidth}
        \centering
        \includegraphics[width=2.8in]{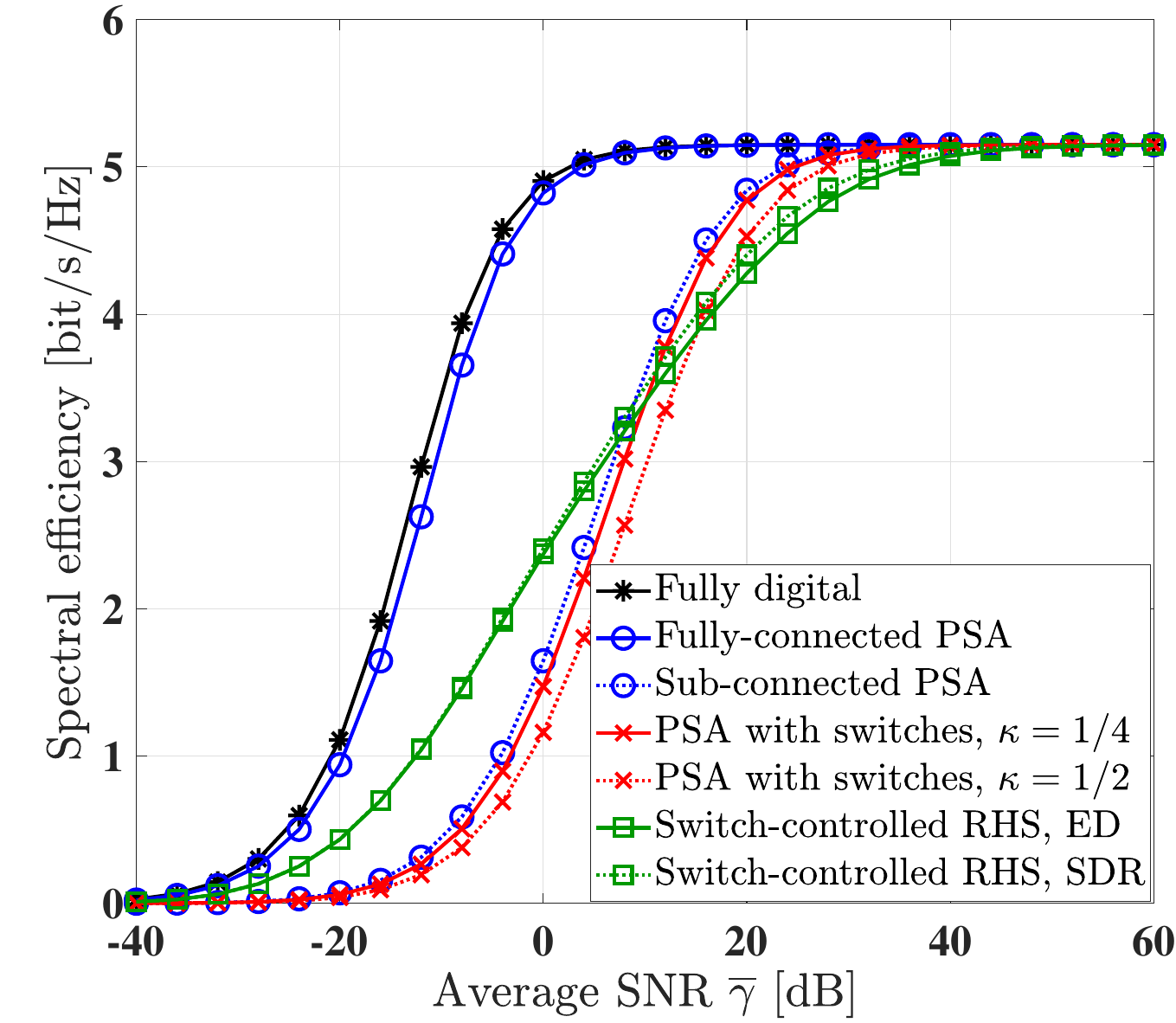}
    \end{minipage}}
    \caption{Comparison of the spectral efficiency versus the average SNR $\overline{\gamma}$ in various beamforming schemes.}\label{Simu_Fig_SE_SNR}
\end{figure}

\begin{figure}[!t]
    \centering
    \subfloat[Hardware quality factors $\varepsilon_\mathrm{t}=\varepsilon_\mathrm{r}=1$.]{\begin{minipage}{1\linewidth}
        \centering
        \includegraphics[width=2.8in]{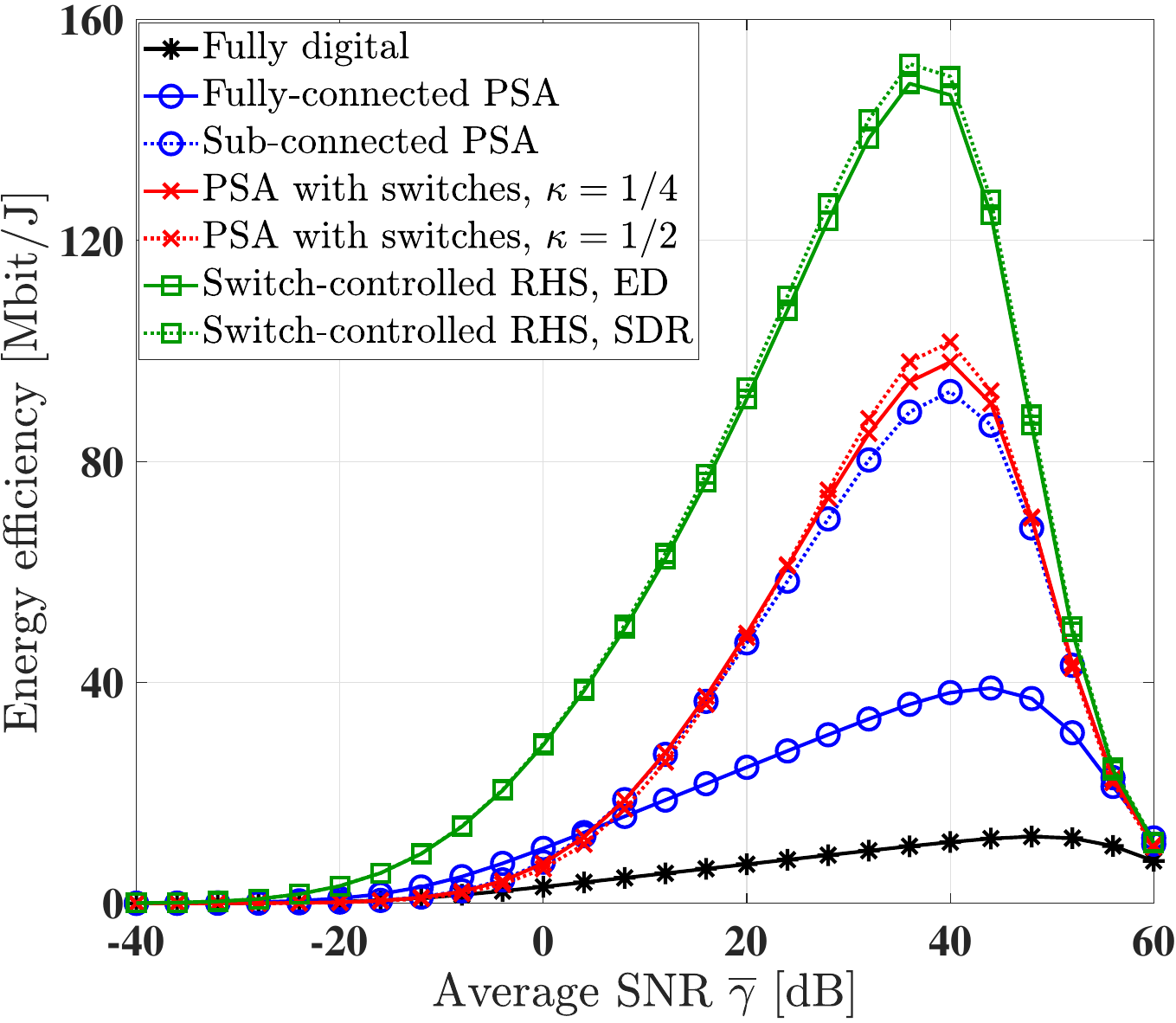}
    \end{minipage}}\\
    \subfloat[Hardware quality factors $\varepsilon_\mathrm{t}=\varepsilon_\mathrm{r}=0.8$.]{\begin{minipage}{1\linewidth}
        \centering
        \includegraphics[width=2.8in]{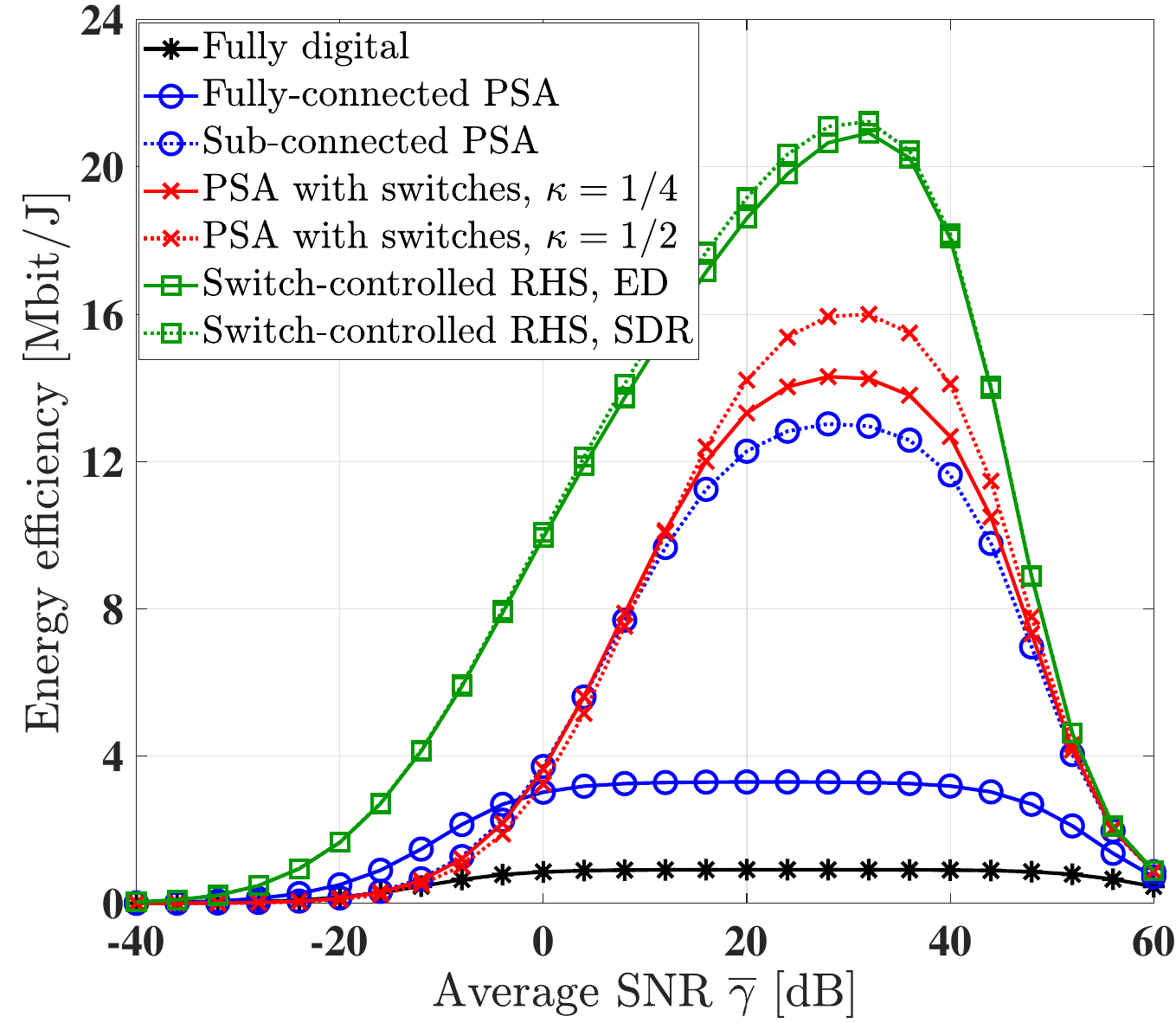}
    \end{minipage}}\\
    \subfloat[Hardware quality factors $\varepsilon_\mathrm{t}=\varepsilon_\mathrm{r}=0.6$.]{\begin{minipage}{1\linewidth}
        \centering
        \includegraphics[width=2.8in]{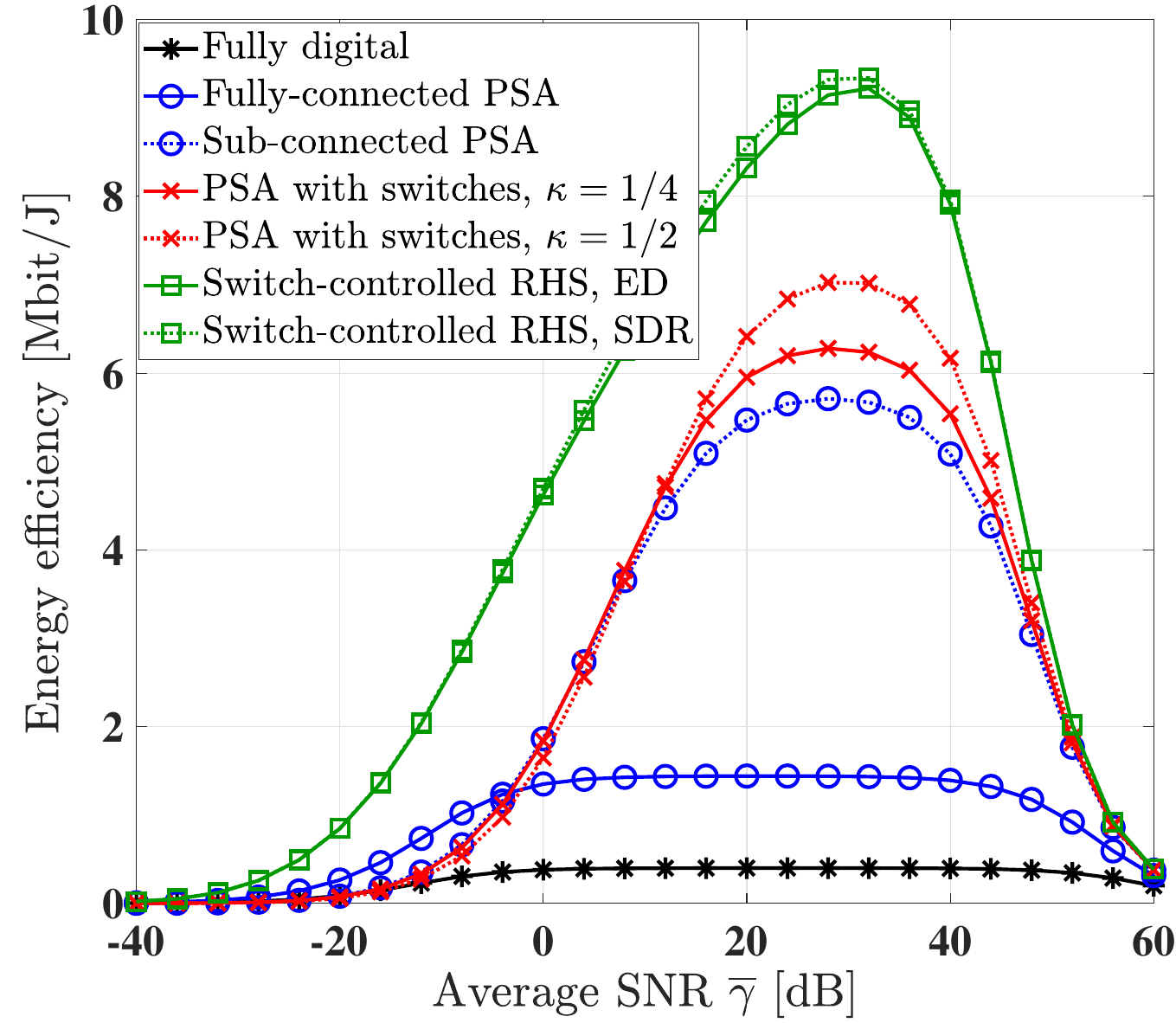}
    \end{minipage}}
    \caption{Comparison of the energy efficiency versus the average SNR $\overline{\gamma}$ in various beamforming schemes.}\label{Simu_Fig_EE_SNR}
\end{figure}

Fig.~\ref{Simu_Fig_SE_SNR} and Fig.~\ref{Simu_Fig_EE_SNR} characterize the spectral efficiency and energy efficiency of different beamforming schemes versus the average SNR $\overline{\gamma}$ associated with different hardware quality factors, respectively. The relationship between the average SNR and the transmit power is $\overline{\gamma}=\frac{\rho\overline{\upsilon}}{\sigma_w^2}$, which means that the transmit power is fixed without being optimized, when the average SNR is given. This shows that the fully digital scheme and the fully-connected PSA scheme achieve better spectral efficiency than the switch-controlled RHS beamformer. This is due to the fact that more RF chains and more phase shifters are deployed in the fully digital scheme and in the fully-connected PSA scheme, respectively. Furthermore, the switch-controlled RHS beamforming scheme has a comparable spectral efficiency to that of the sub-connected PSA and the PSA with switches. By contrast, in terms of the energy efficiency, the switch-controlled RHS beamformer has the lowest power consumption. Next comes the PSA associated with switches. Fig.~\ref{Simu_Fig_SE_SNR} and Fig.~\ref{Simu_Fig_EE_SNR} show that in the switch-controlled RHS beamformer, the ED method almost has the same performance to the SDR optimization method, while at much lower complexity. Furthermore, it shows that when the transceivers have imperfect hardware quality, i,e. $\varepsilon_\mathrm{t}<1$ and $\varepsilon_\mathrm{r}<1$, the spectral efficiency is limited in the high-SNR region by the hardware impairments and saturates at a constant value of $K\log_2\Big(1+\frac{\varepsilon_\mathrm{r}\varepsilon_\mathrm{t}}
{1-\varepsilon_\mathrm{r}\varepsilon_\mathrm{t}}\Big)=11.79$ bit/s/Hz when $\varepsilon_\mathrm{t}=\varepsilon_\mathrm{r}=0.8$ and at $K\log_2\Big(1+\frac{\varepsilon_\mathrm{r}\varepsilon_\mathrm{t}}
{1-\varepsilon_\mathrm{r}\varepsilon_\mathrm{t}}\Big)=5.151$ bit/s/Hz when $\varepsilon_\mathrm{t}=\varepsilon_\mathrm{r}=0.6$, which agrees with our analysis in (\ref{Theoretical_Analysis_1}). Moreover, for the switch-controlled holographic beamforming architecture, according to (\ref{Theoretical_Analysis_4}), the energy efficiency upper bound is $\frac{BK\log_2\big(1+\frac{\varepsilon_\mathrm{r}\varepsilon_\mathrm{t}}
{1-\varepsilon_\mathrm{r}\varepsilon_\mathrm{t}}\big)}
{P_{\mathrm{Syn}}+MP_{\mathrm{RF}}+NP_{\mathrm{SW}}}=22.99$ Mbit/J when $\varepsilon_\mathrm{t}=\varepsilon_\mathrm{r}=0.8$ and $\frac{BK\log_2\big(1+\frac{\varepsilon_\mathrm{r}\varepsilon_\mathrm{t}}
{1-\varepsilon_\mathrm{r}\varepsilon_\mathrm{t}}\big)}
{P_{\mathrm{Syn}}+MP_{\mathrm{RF}}+NP_{\mathrm{SW}}}=10.04$ Mbit/J when $\varepsilon_\mathrm{t}=\varepsilon_\mathrm{r}=0.6$.

\begin{figure}[t]
    \centering
    \includegraphics[width=2.8in]{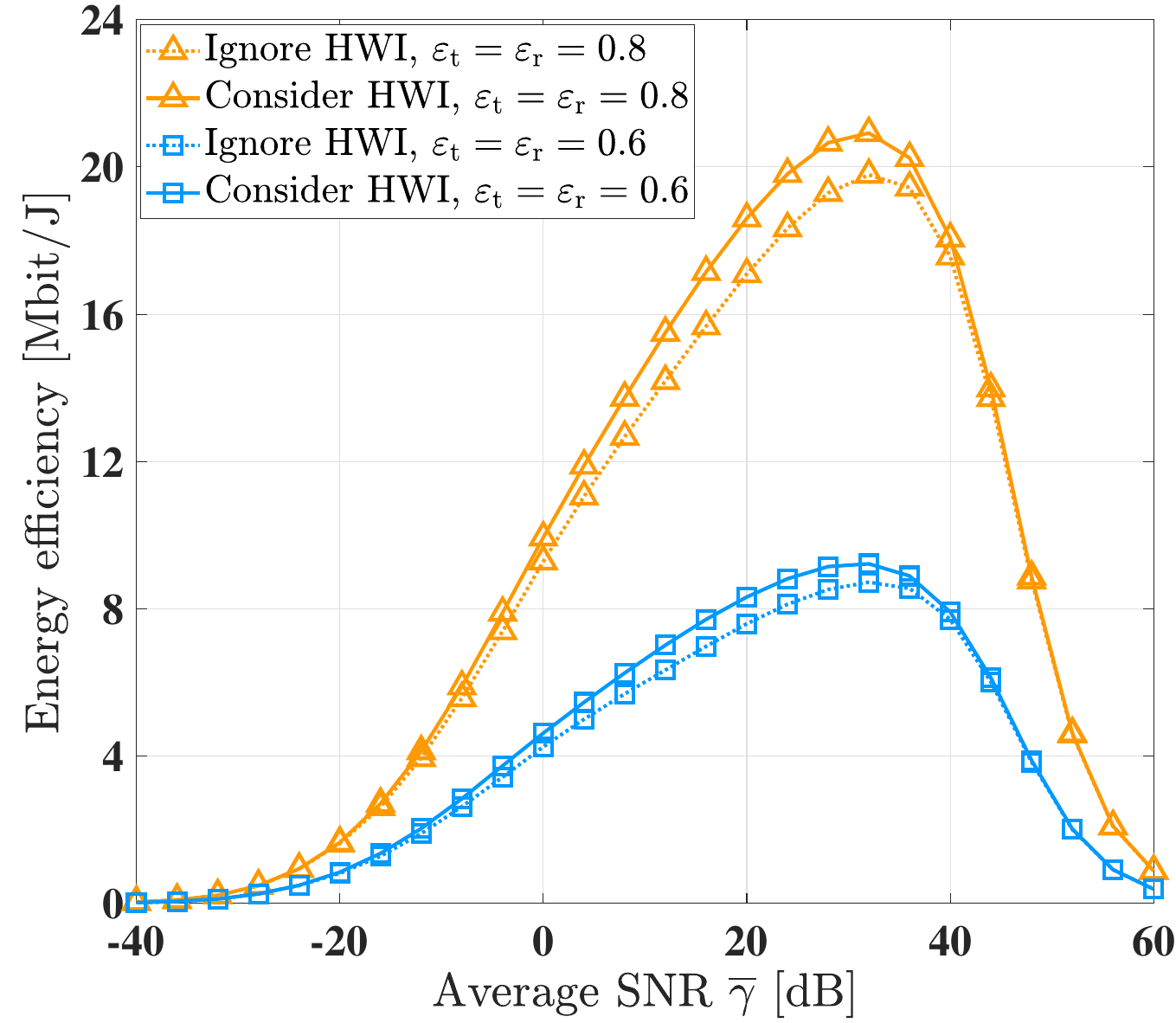}
    \caption{Comparison of the energy efficiency versus the average SNR $\overline{\gamma}$ in the switch-controlled RHS-based beamforming scheme, with the effect of HWI is considered or ignored in the beamforming algorithms.}\label{Simu_Fig_EE_SNR_com}
\end{figure}

Fig.~\ref{Simu_Fig_EE_SNR_com} shows the energy efficiency of the switch-controlled RHS-based beamformers versus the average SNR along with the hardware quality factors of $\varepsilon_\mathrm{t}=\varepsilon_\mathrm{r}=0.8$ and $\varepsilon_\mathrm{t}=\varepsilon_\mathrm{r}=0.6$, respectively. The legend `\textit{Ignore HWI}' means that the HWIs are ignored in the power sharing and the state-of-the-art water-filling method is employed. By contrast, the legend `\textit{Consider HWI}' means that the HWIs are considered in the power sharing and Algorithm~\ref{algorithm_1} is employed. This shows that increased energy efficiency can be attained when the effect of HWIs is taken into account. Specifically, the maximum energy efficiency in the switch-controlled RHS beamformer scheme can be achieved when the transmit power is $\rho=32\mathrm{dB}$. When the hardware quality is imperfect as indicated by the factor of $\varepsilon_\mathrm{t}=\varepsilon_\mathrm{r}=0.6$ and the HWIs are ignored in the power sharing, approximately 8.7 Mbit/J and 8.9 Mbit/J of maximum energy efficiency can be achieved by the ED method and by the SDR method, respectively. By contrast, when the HWIs are explicitly considered in the power sharing design, approximately 9.2 Mbit/J and 9.3 Mbit/J of maximum energy efficiency can be achieved by the ED method and by the SDR method, respectively. This means that about 1.5 Mbit/J energy efficiency gain can be achieved by considering the HWIs in the power sharing design.

\begin{figure}[!t]
    \centering
    \subfloat[Hardware quality factors $\varepsilon_\mathrm{t}=\varepsilon_\mathrm{r}=1$.]{\begin{minipage}{1\linewidth}
        \centering
        \includegraphics[width=2.8in]{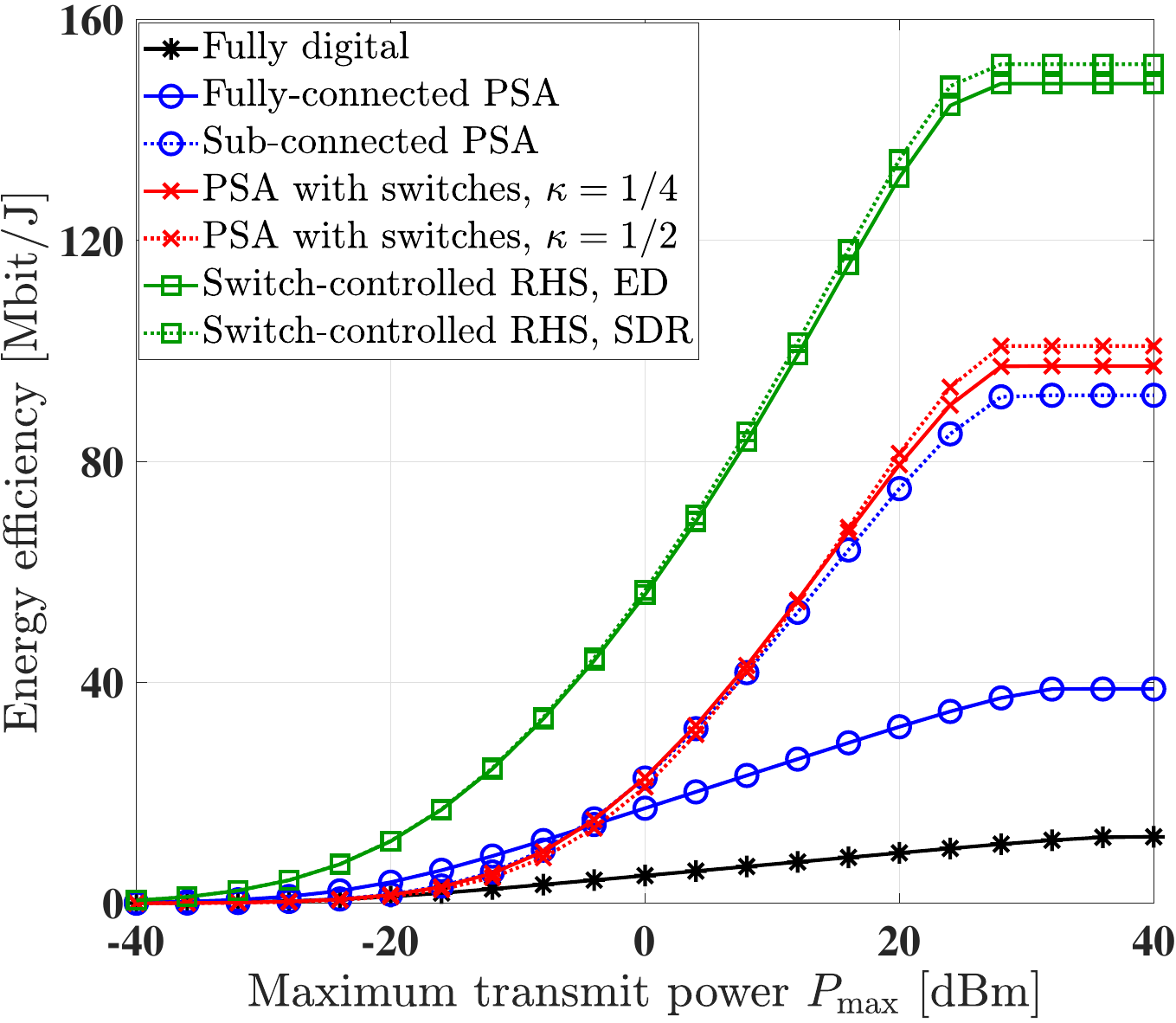}
    \end{minipage}}\\
    \subfloat[Hardware quality factors $\varepsilon_\mathrm{t}=\varepsilon_\mathrm{r}=0.8$.]{\begin{minipage}{1\linewidth}
        \centering
        \includegraphics[width=2.8in]{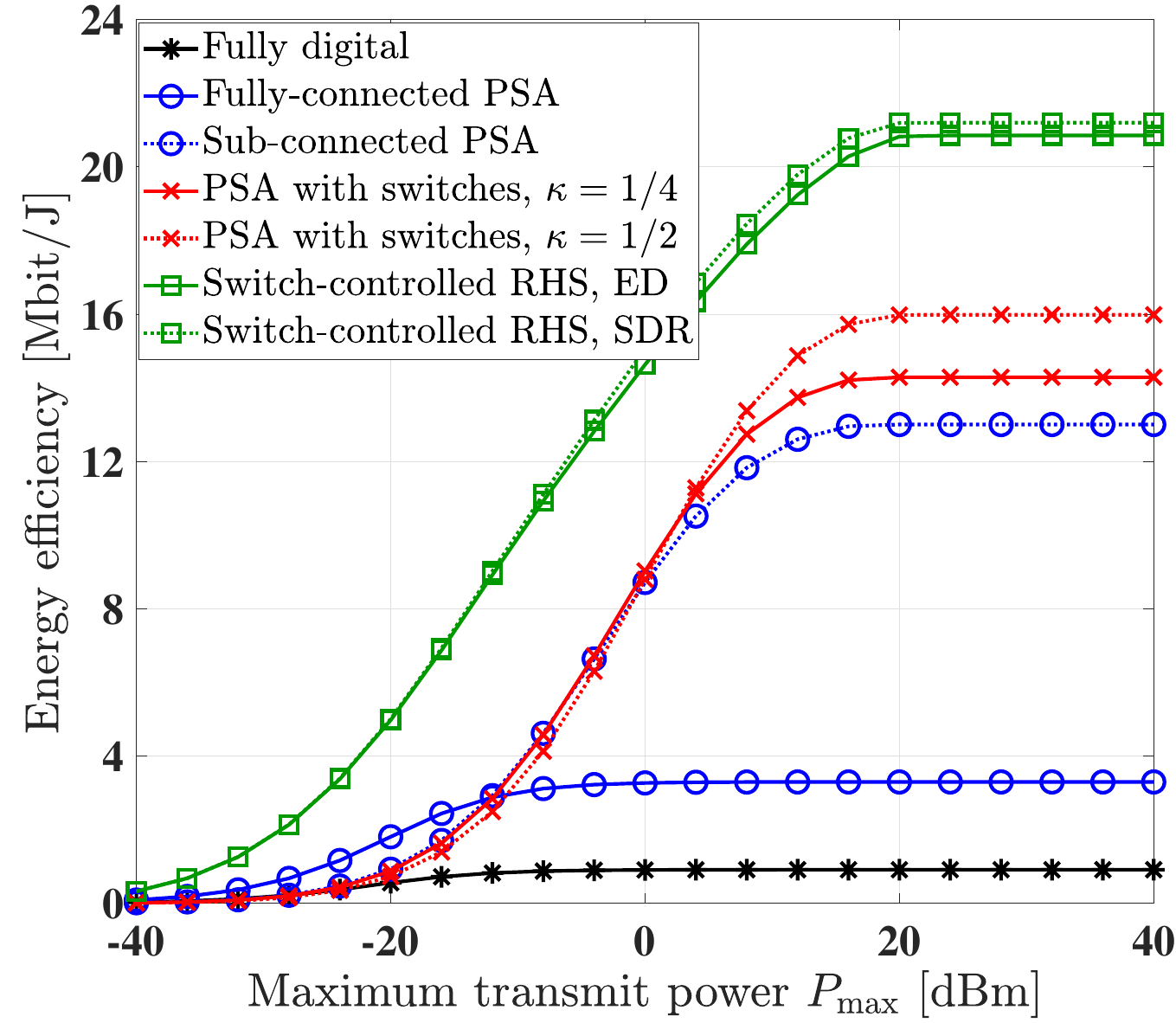}
    \end{minipage}}\\
    \subfloat[Hardware quality factors $\varepsilon_\mathrm{t}=\varepsilon_\mathrm{r}=0.6$.]{\begin{minipage}{1\linewidth}
        \centering
        \includegraphics[width=2.8in]{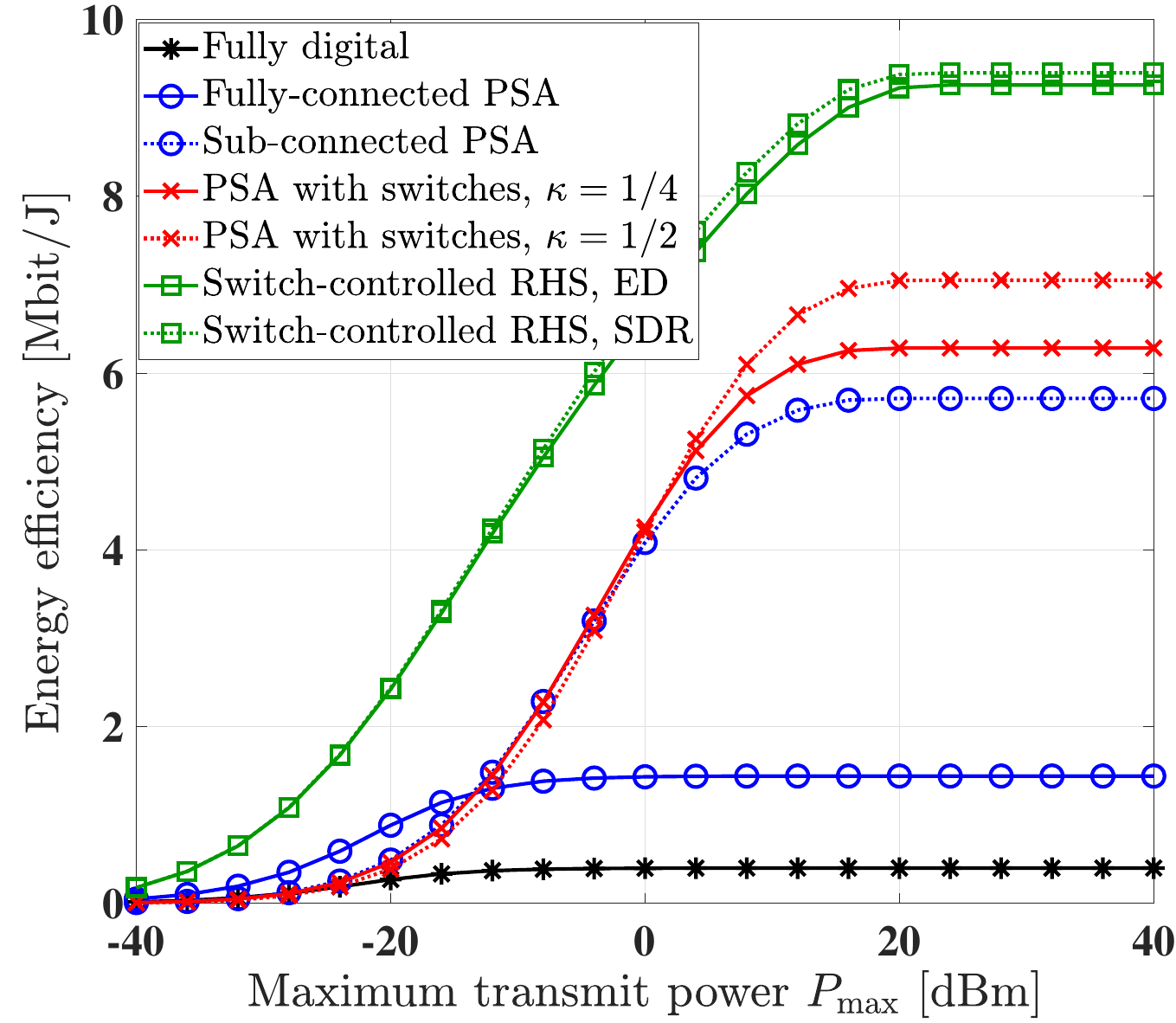}
    \end{minipage}}
    \caption{Comparison of the energy efficiency versus the maximum transmit power budget $P_\mathrm{max}$ in various beamforming schemes.}\label{Simu_Fig_EE_Pt}
\end{figure}

Fig.~\ref{Simu_Fig_EE_Pt} compares the energy efficiency versus the maximum transmit power budget $P_\mathrm{max}$ of various beamformers. Fig.~\ref{Simu_Fig_EE_Pt} (a) shows that when the hardware quality is ideal, the switch-controlled RHS beamformer outperforms the benchmarks. Furthermore, the energy efficiency saturates when the maximum transmit power budget is $P_\mathrm{max}\approx30\mathrm{dBm}$, which means that the optimal energy efficiency is achieved when $P_\mathrm{max}\approx30\mathrm{dBm}$. Fig.~\ref{Simu_Fig_EE_Pt} (b) and Fig.~\ref{Simu_Fig_EE_Pt} (c) show that the switch-controlled RHS beamformer performs constantly better than other beamforming schemes, regardless of the hardware quality.

\begin{figure}[t]
    \centering
    \includegraphics[width=2.8in]{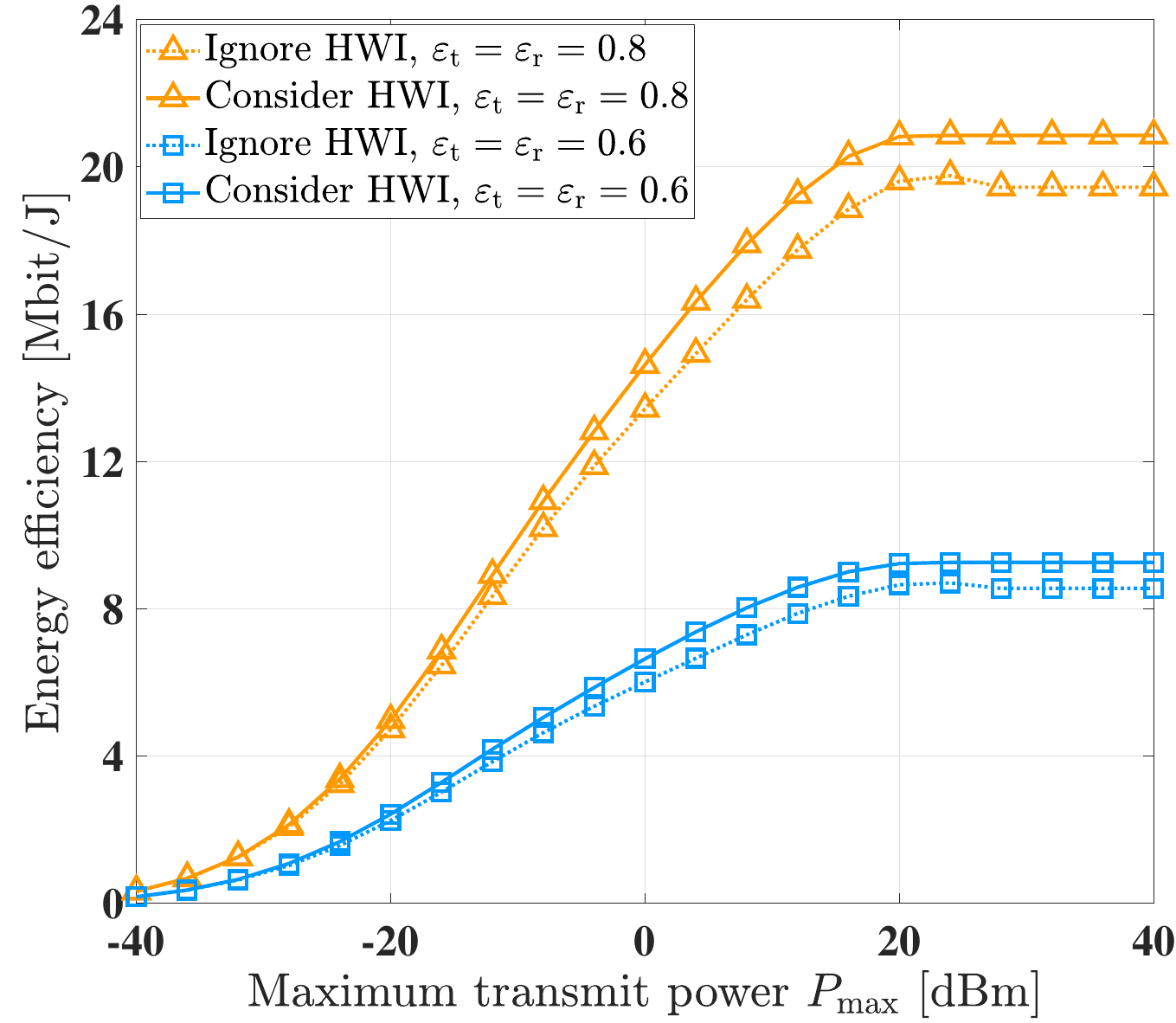}
    \caption{Comparison of the energy efficiency versus the maximum transmit power budget $P_\mathrm{max}$ in the RHS-based beamforming scheme, with the effect of HWI is considered or ignored in the beamforming algorithms.}\label{Simu_Fig_EE_Pt_com}
\end{figure}

Fig.~\ref{Simu_Fig_EE_Pt_com} compares the energy efficiency versus the maximum transmit power budget $P_\mathrm{max}$ of the switch-controlled RHS-based beamformer, when the effect of HWIs is either considered or ignored in the beamformer design. Explicitly, considering the effect of HWIs can bring about further energy efficiency enhancements. Furthermore, when $P_\mathrm{max}>20\mathrm{dBm}$, the energy efficiency actually decays upon increasing the maximum transmit power budget if the deleterious effect of HWIs is ignored in the power sharing design. This is due to the fact that the effect of HWIs becomes more grave upon increasing the transmit power. Moreover, Fig.~\ref{Simu_Fig_EE_Pt_com} shows that when $P_\mathrm{max}>20\mathrm{dBm}$, considering the HWIs in the power sharing design can bring about 1.4 Mbit/J and 1.7 Mbit/J energy efficiency performance gain, when the hardware quality factors are $\varepsilon_\mathrm{t}=\varepsilon_\mathrm{r}=0.8$ and $\varepsilon_\mathrm{t}=\varepsilon_\mathrm{r}=0.6$, respectively.

\begin{figure}[t]
    \centering
    \includegraphics[width=2.8in]{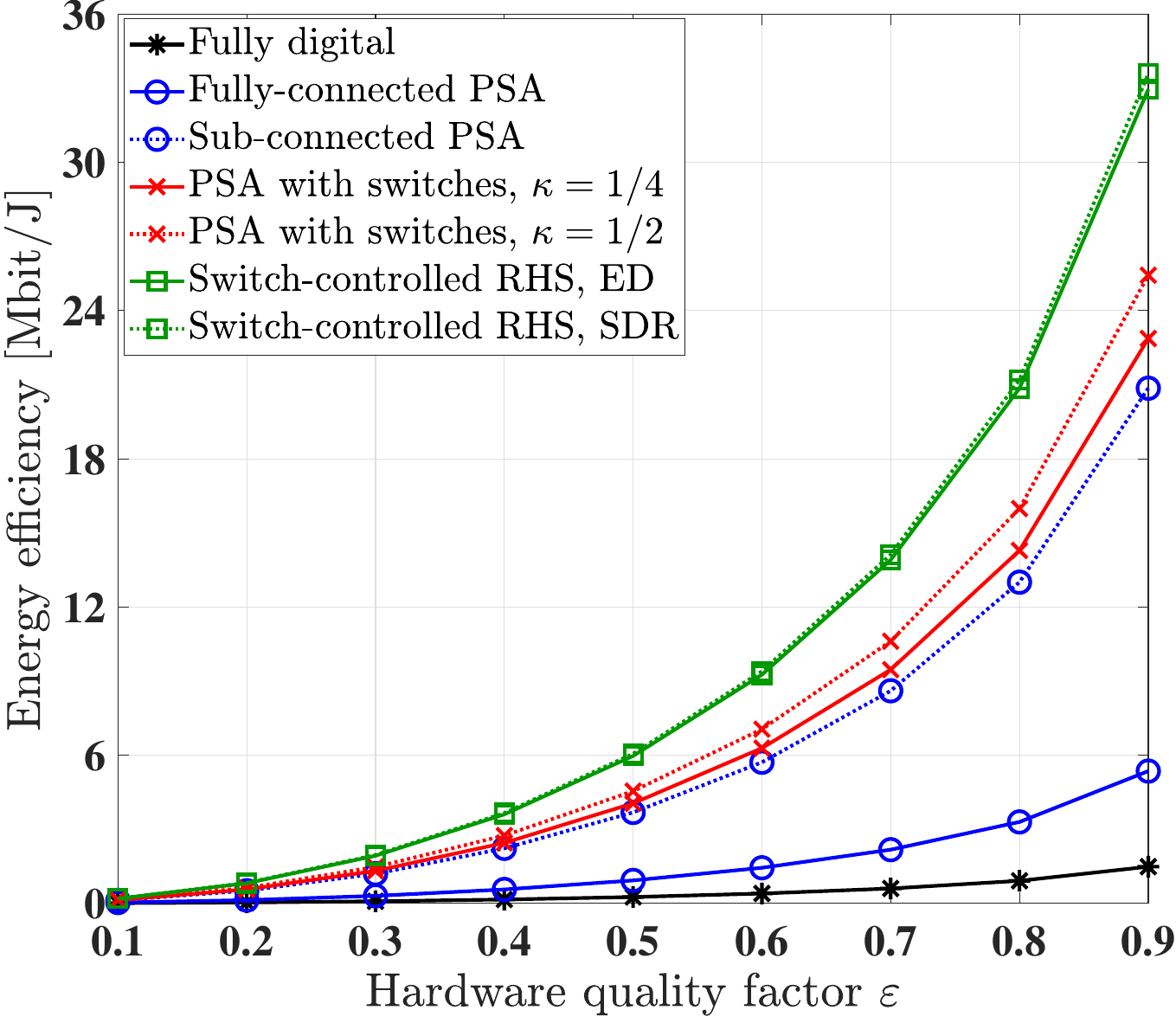}
    \caption{Comparison of the energy efficiency versus the hardware quality factors in various beamforming schemes, where $\varepsilon=\varepsilon_\mathrm{t}=\varepsilon_\mathrm{r}$.}\label{Simu_Fig_EE_epsilon}
\end{figure}

Fig.~\ref{Simu_Fig_EE_epsilon} compares the energy efficiency versus the hardware quality factors of various beamformers, where $\varepsilon=\varepsilon_\mathrm{t}=\varepsilon_\mathrm{r}$. This shows that the energy efficiency increases with the improvement of the hardware quality factors. Moreover, the switch-controlled RHS beamformer consistently outperforms the other schemes.

\begin{figure}[t]
    \centering
    \includegraphics[width=2.8in]{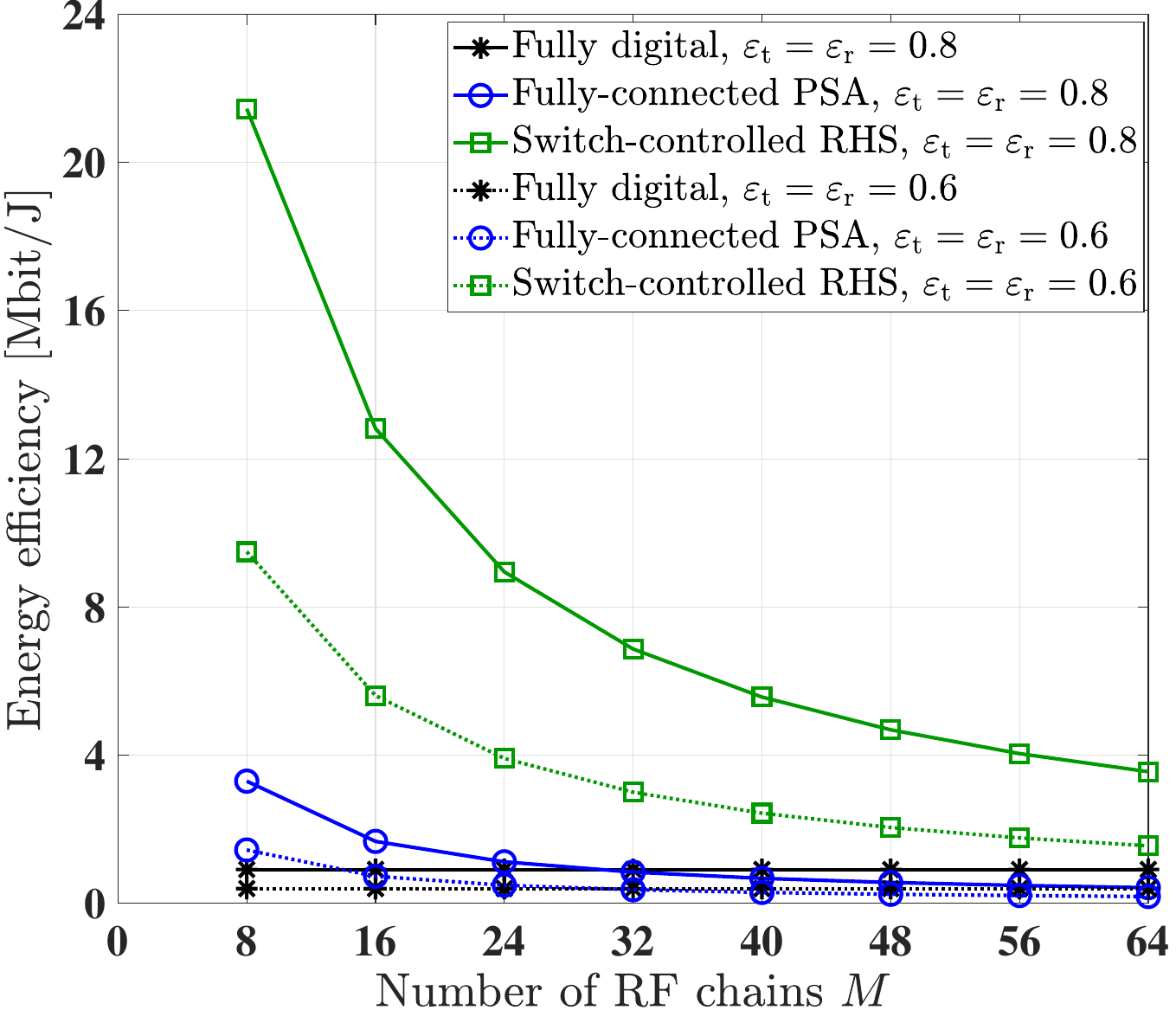}
    \caption{Comparison of the energy efficiency versus the number of RF chains for various beamforming schemes.}\label{Simu_Fig_EE_M}
\end{figure}

\begin{figure}[t]
    \centering
    \includegraphics[width=2.8in]{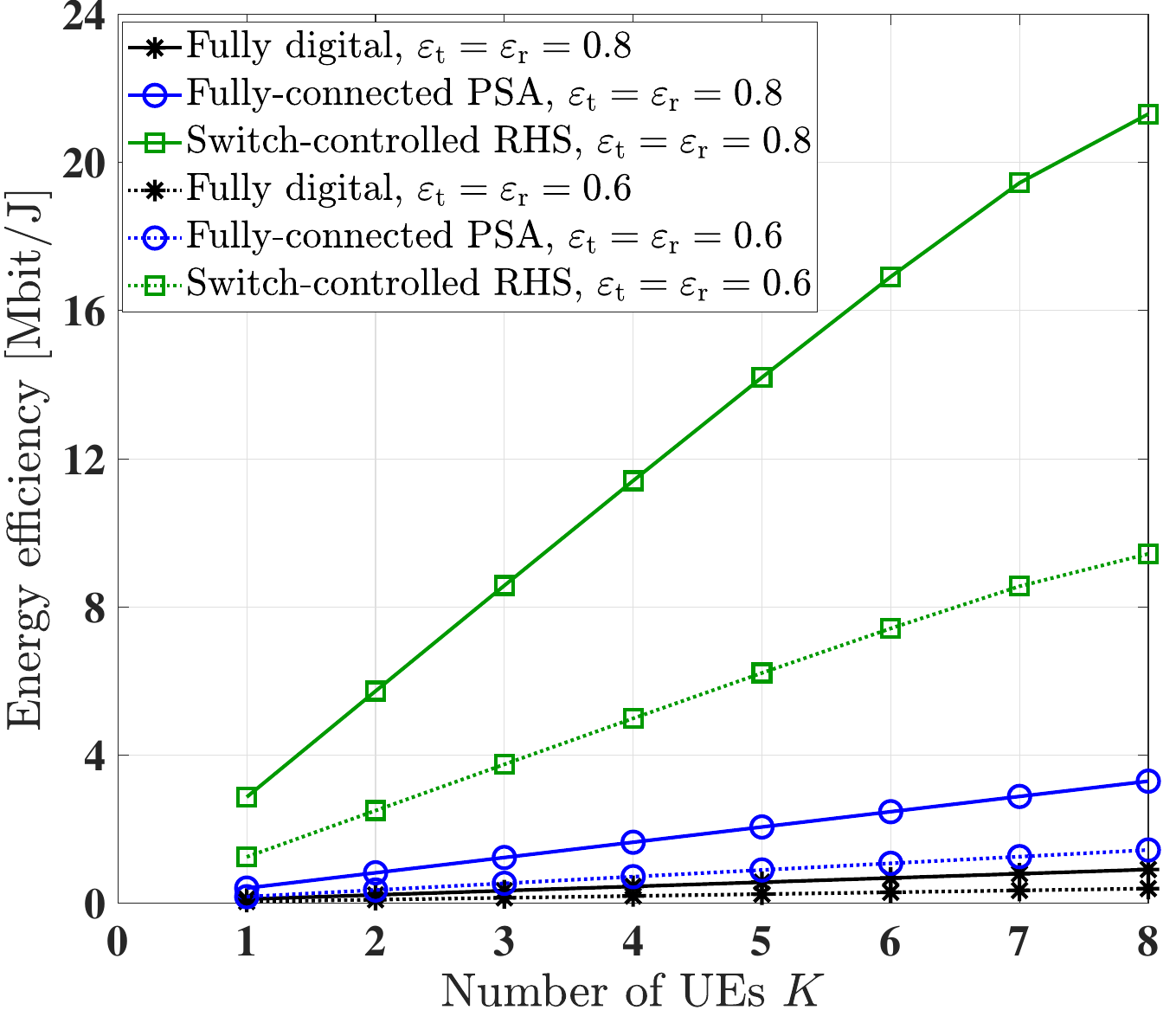}
    \caption{Comparison of the energy efficiency versus the number of UEs for various beamforming schemes.}\label{Simu_Fig_EE_K}
\end{figure}

\begin{figure}[t]
    \centering
    \includegraphics[width=2.8in]{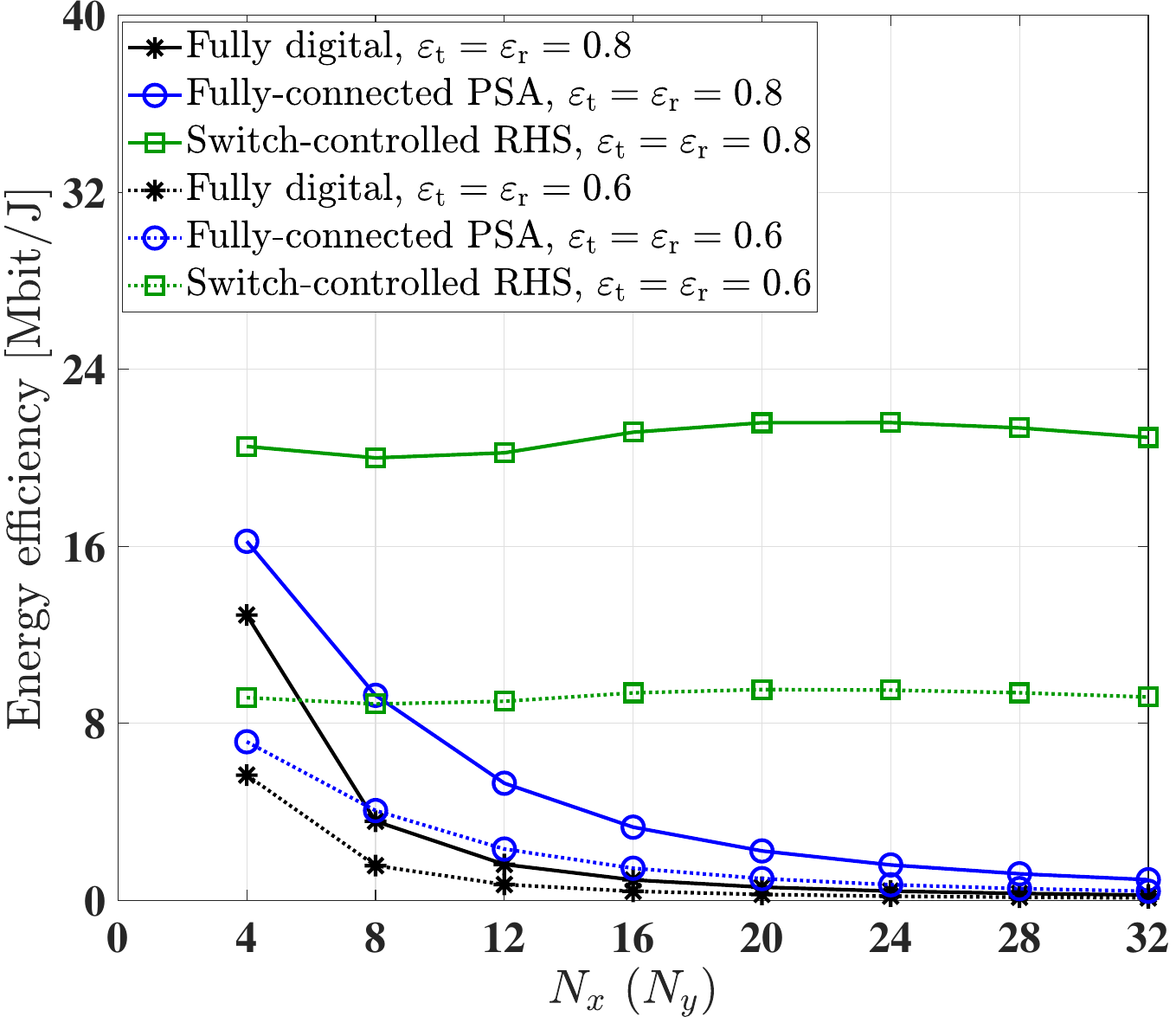}
    \caption{Comparison of the energy efficiency versus the number of RHS elements for various beamforming schemes.}\label{Simu_Fig_EE_N}
\end{figure}

Fig.~\ref{Simu_Fig_EE_M}, Fig.~\ref{Simu_Fig_EE_K} and Fig.~\ref{Simu_Fig_EE_N} compare the energy efficiency versus the number of RF chains, number of UEs and number of RHS elements in various beamformers, respectively. Firstly, Fig.~\ref{Simu_Fig_EE_M} shows that the energy efficiency decreases upon deploying more RF chains due to the power consumption of RF chains. Furthermore, the switch-controlled RHS beamforming architectures have higher energy efficiency than the fully digital beamforming architecture and fully-connected PSA beamforming architecture. Fig.~\ref{Simu_Fig_EE_K} portrays that the energy efficiency increases almost linearly with the number of UEs. Fig.~\ref{Simu_Fig_EE_N} shows that with the increase of the number of RHS elements, the energy efficiency of the switch-controlled RHS beamforming architecture is much higher than that of the fully digital beamforming architecture and fully-connected PSA beamforming architecture. Furthermore, the energy efficiency of the fully digital beamforming architecture, the fully-connected PSA beamforming architecture was reduced upon increasing the number of RHS elements due to the high power consumption of the RF chains and phase shifters. By contrast, the energy efficiency of the switch-controlled RHS beamforming architecture remains almost unchanged, as the number of RHS elements increases due to the lower power consumption of the switches.

\begin{figure}[!t]
    \centering
    \subfloat[Hardware quality factors $\varepsilon_\mathrm{t}=\varepsilon_\mathrm{r}=1$.]{\begin{minipage}{1\linewidth}
        \centering
        \includegraphics[width=2.8in]{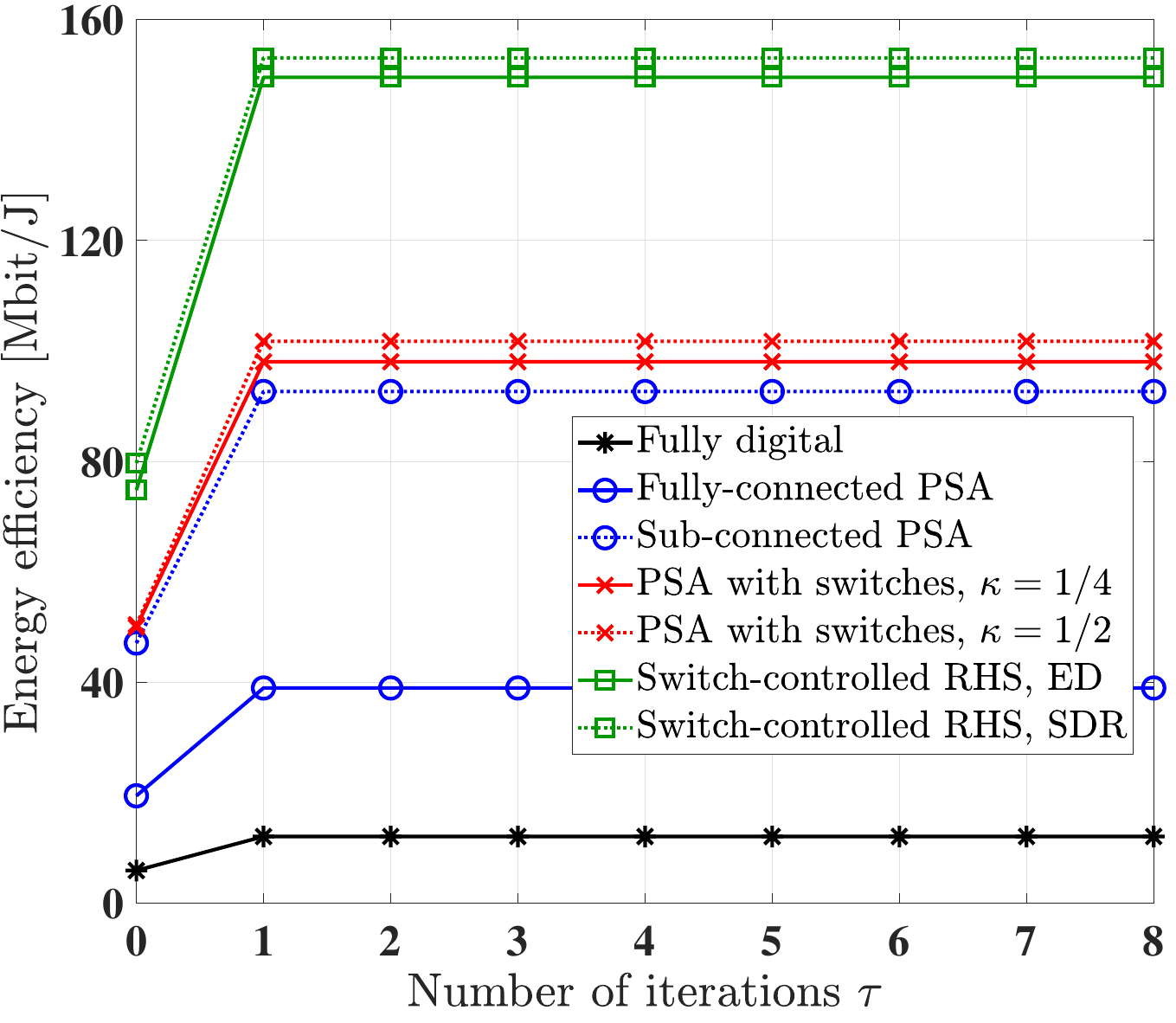}
    \end{minipage}}\\
    \subfloat[Hardware quality factors $\varepsilon_\mathrm{t}=\varepsilon_\mathrm{r}=0.8$.]{\begin{minipage}{1\linewidth}
        \centering
        \includegraphics[width=2.8in]{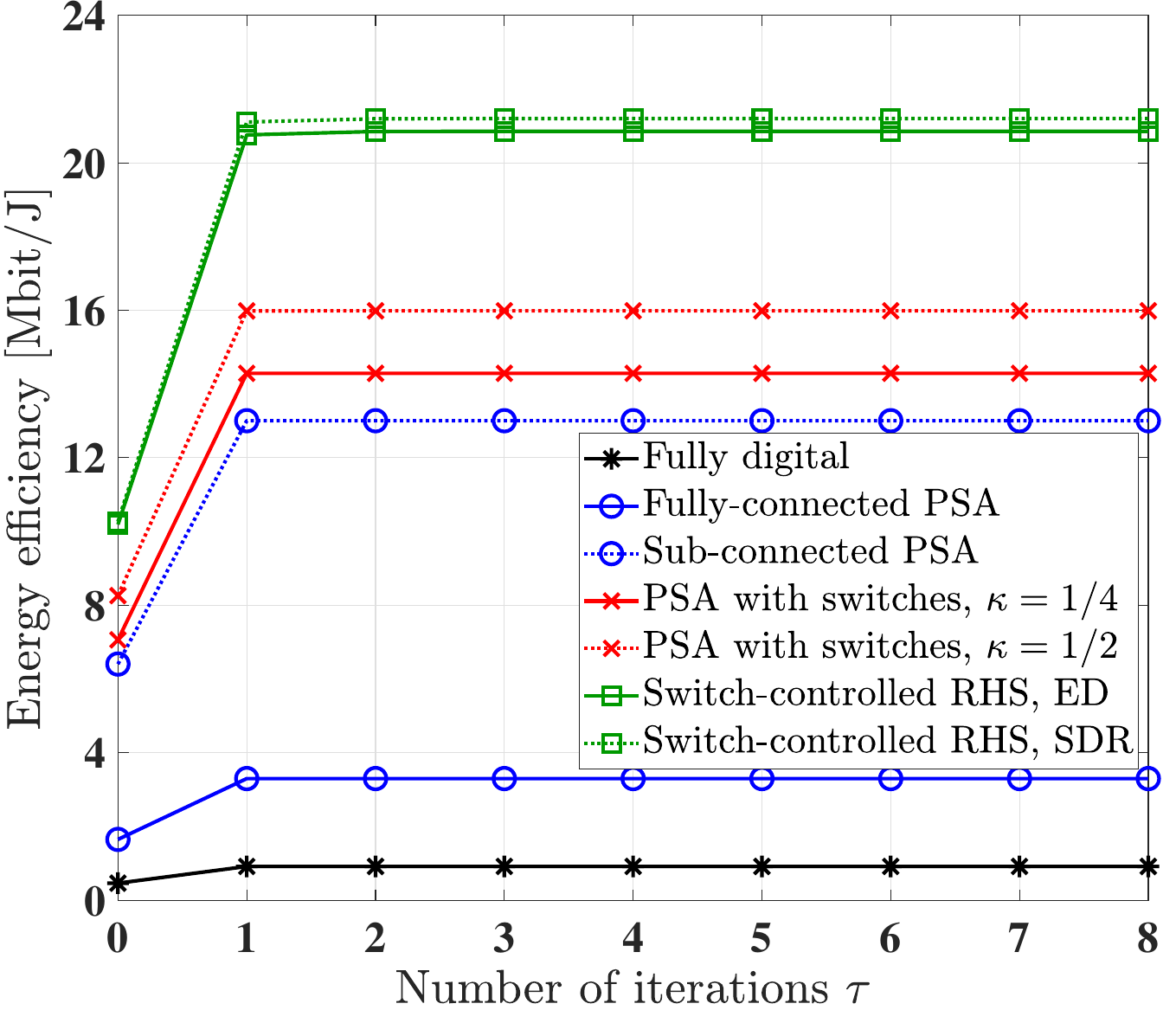}
    \end{minipage}}\\
    \subfloat[Hardware quality factors $\varepsilon_\mathrm{t}=\varepsilon_\mathrm{r}=0.6$.]{\begin{minipage}{1\linewidth}
        \centering
        \includegraphics[width=2.8in]{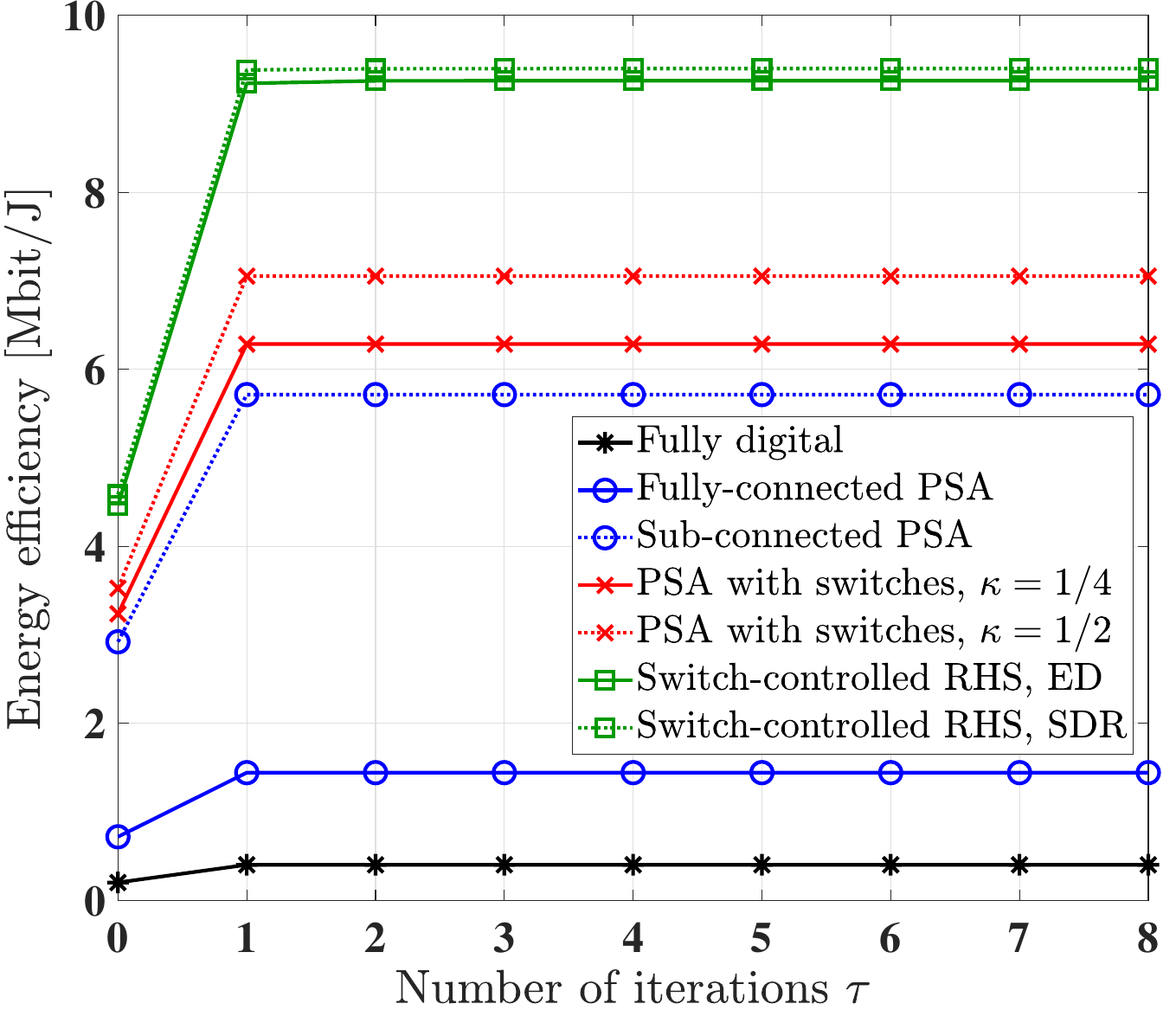}
    \end{minipage}}
    \caption{Comparison of the energy efficiency versus the number of iterations for various beamforming schemes.}\label{Simu_Fig_EE_Iteration}
\end{figure}

In Fig.~\ref{Simu_Fig_EE_Iteration}, the energy efficiency versus the number of iterations $\tau$ used by the alternating optimization algorithm is presented for the different hardware quality factors. It shows that the convergence speed in our proposed alternating optimization algorithm is excellent. Furthermore, the alternating optimization algorithm in the switch-controlled RHS-aided beamforming architecture converges when $\tau=2$.

\section{Conclusions}\label{Conclusion}
The problem of maximizing the energy efficiency for switch-controlled RHS beamforming architectures was formulated by optimizing the digital beamformer, holographic beamformer, the total transmit power and the power sharing ratio of users, while considering the effect of HWIs. Our simulation results demonstrated that the switch-controlled RHS beamformers outperform the conventional fully digital beamformer and hybrid beamformer based on the PSA scheme in terms of the energy efficiency attained. The energy-efficient RHS beamforming architecture constitutes a promising option for the practical holographic MIMO implementation.

\bibliographystyle{IEEEtran}
\bibliography{IEEEabrv,TAMS}
\end{document}